\documentclass[a4paper,11pt]{article}
\pdfoutput=1 

\usepackage{jcappub} 

\usepackage[T1]{fontenc} 

\usepackage[normalem]{ulem}

\usepackage[utf8]{inputenc}
\usepackage{aas_macros}
\usepackage{amsfonts}
\usepackage{amsmath}
\usepackage{amssymb}
\usepackage{bm}
\usepackage{graphicx}
\usepackage[dvipsnames]{xcolor}
\usepackage{hyperref}
\hypersetup{colorlinks=true,allcolors=teal}

\defcitealias{ps22}{PS22}
\defcitealias{ps23}{PS23}
\defcitealias{ps25a}{PS25a}
\defcitealias{ps25b}{PS25b}
\defcitealias{ps26a}{PS26}

\newcommand{\p}{\ensuremath{\partial}}

\newcommand{\yy}{\ensuremath{\mathbf{y}}}

\renewcommand{\aap}[1]{\ensuremath{\mathbf{a}_{\rm (p)#1}}}

\newcommand{\Mh}{\ensuremath{h^{-1}M_{\odot}}}

\newcommand{\Mpch}{\ensuremath{h^{-1}{\rm Mpc}}}
\newcommand{\Gpch}{\ensuremath{h^{-1}{\rm Gpc}}}
\newcommand{\hMpc}{\ensuremath{h\,{\rm Mpc}^{-1}}}

\newcommand{\kms}{\ensuremath{{\rm km\,s}^{-1}}}

\newcommand{\der}{\ensuremath{{\rm d}}}

\newcommand{\eqn}[1]{equation~\eqref{#1}}
\newcommand{\eqns}[1]{equations~\eqref{#1}}
\newcommand{\ph}[1]{\phantom{#1}}

\newcommand{\beq}{\begin{equation}}
\newcommand{\eeq}{\end{equation}}
\newcommand{\Cal}[1]{\ensuremath{\mathcal{#1}}}

\newcommand{\xiell}[1]{\ensuremath{\xi^{(#1)}}}
\newcommand{\xiellobs}[1]{\ensuremath{\xi_{\rm obs}^{(#1)}}}
\newcommand{\xiellsim}[1]{\ensuremath{\hat \xi^{(#1)}_{\rm obs}}}
\newcommand{\Sigell}[1]{\ensuremath{\Sigma^{(#1)2}}}
\newcommand{\Sigellsim}[1]{\ensuremath{\hat\Sigma_{\rm obs}^{(#1)2}}}
\newcommand{\Sigellobs}[1]{\ensuremath{\Sigma_{\rm obs}^{(#1)2}}}
\newcommand{\Dellsq}[1]{\ensuremath{\Delta^{(#1)2}}}
\newcommand{\Dellsqobs}[1]{\ensuremath{\Delta^{(#1)2}_{\rm obs}}}
\newcommand{\sigv}{\ensuremath{\sigma_{\rm v}}}
\newcommand{\Pell}[1]{\ensuremath{\mathcal{P}_{#1}}}
\newcommand{\tjred}[3]{\ensuremath{\begin{pmatrix}
#1 & #2 & #3 \\
0 & 0 & 0 
\end{pmatrix}}}
\newcommand{\tjs}[6]{\ensuremath{\left(\begin{smallmatrix}
#1 & #2 & #3 \\
#4 & #5 & #6 
\end{smallmatrix}\right)}}

\renewcommand{\ss}{\ensuremath{\mathbf{s}}}

\newcommand{\qbar}[2]{\ensuremath{\bar{q}^{(#1)}_{#2}}}

\newcommand{\aiso}{\ensuremath{\alpha_{\rm iso}}}
\newcommand{\aAP}{\ensuremath{\alpha_{\rm AP}}}
\newcommand{\aperp}{\ensuremath{\alpha_\perp}}
\newcommand{\nbarv}[1]{\ensuremath{\bar n^{(#1)}_{\rm v}}}
\newcommand{\Daiso}{\ensuremath{\Delta\alpha_{\rm iso}}}
\newcommand{\DaAP}{\ensuremath{\Delta\alpha_{\rm AP}}}

\newcommand{\biseq}{\texttt{BiSequential}}
\newcommand{\abacussummit}{\texttt{AbacusSummit}}
\newcommand{\base}{\texttt{c000}}
\newcommand{\desi}{\texttt{DESI-LRG2}}
\newcommand{\euclid}{\texttt{Euclid-ELG}}

\title{\boldmath Impact of fiducial cosmology in model-agnostic cosmological inference with the BAO feature}

\author[a]{Aseem Paranjape}
\author[b,c]{and Ravi K. Sheth}

\affiliation[a]{Inter-University Centre for Astronomy \& Astrophysics,\\ Ganeshkhind, Post Bag 4, Pune 411007, India}
\affiliation[b]{Center for Particle Cosmology, University of Pennsylvania,\\ 209 S. 33rd St., Philadelphia, PA 19104, USA}

\affiliation[c]{The Abdus Salam International Center for Theoretical Physics,\\ Strada Costiera, 11, Trieste 34151, Italy}
 
\emailAdd{aseem@iucaa.in}
\emailAdd{shethrk@physics.upenn.edu}

\abstract{
In recent work, we have developed a model-agnostic framework for cosmological inference using the baryon acoustic oscillation (BAO) feature in redshift space. 
The framework, which we dub the `Zel'dovich smearing approximation', involves several ingredients, including an optimal basis for the linear theory correlation function in real space, a Gaussian kernel that describes the smearing of the BAO feature due to bulk flows, and a model for the impact of scale dependent bias and mode coupling in redshift space. 
Unlike traditional, template-based approaches, this framework does not assume any particular cosmological model. 
An important ingredient -- which impacts both the traditional as well as model-agnostic frameworks -- is the geometric distortion introduced by the inevitable choice of the fiducial cosmology used for converting observed angles and redshifts to comoving separations. 
In this work we show how this distortion, which was not yet included in our framework, affects the output of the Zel'dovich smearing approximation. 
We test our completed framework on two halo samples derived from the \abacussummit\ simulation suite and designed to mimic galaxy samples from the ongoing DESI and \emph{Euclid} surveys, respectively. 
We show that our framework produces unbiased cosmological constraints when applied to these semi-realistic samples and compare the organization of cosmological information in the model-agnostic and traditional approaches. 
We end with a roadmap to a detailed comparison of the performance of the traditional approach with the Zel'dovich smearing approximation, which is `data ready' as presented here. 
}


\begin{document}
\maketitle
\flushbottom

\section{Introduction}
\label{sec:intro}
The large scale structure of the Universe contains clues that can potentially answer some of the most challenging, outstanding fundamental questions in cosmology, such as the nature of gravity and the cause of the late-time acceleration of the cosmic expansion, among others. The baryon acoustic oscillation (BAO) feature in the clustering of cosmic tracers such as galaxies, quasars and the Lyman-$\alpha$ forest has emerged as one of the most robust and informative probes that can access this cosmological information \cite{eisenstein+05}. Its inherently 3-dimensional nature in redshift space, combined with the large lever arm provided by tracer samples spread over a range of cosmic epochs, has made the BAO feature a primary science driver in a number of ongoing and upcoming surveys such as those by the Dark Energy Spectroscopic Instrument (DESI; \cite{DESI}) collaboration, the \emph{Euclid} mission \cite{euclid25-missionoverview}, the 4MOST survey \cite{dejong+19-4MOST}, etc.

Although the BAO feature is a clean prediction of linear theory in the standard cosmological model, its actual location, shape and amplitude in statistics like the 2-point correlation function (2pcf) of a tracer sample is expected to be affected by a number of contaminants related to nonlinear gravitational evolution and the astrophysics defining the tracers themselves \cite{cs08}. The promise of the BAO feature as a cosmological probe arises from the expectation that, due to the cosmological size of the feature ($\sim100\Mpch$ comoving) it should be possible to model these contaminating effects using a relatively small number of nuisance parameters, leaving the cosmologically interesting information largely untouched. To what extent this can be achieved in real data has been the focus of intense research over the last two decades \cite{eisenstein+07,pw08,pwc09,padmanabhan+12,kirkby+13,bph15,damico+20,isz20,shapefit2021,fullShapeXir,chen+24,perez-fernandez+25,nonPTrsd,novell-masot+25-DESIPkBk,bst24,bst25,asensio-rivera+26,bayer+26}. 

Traditional approaches to maximizing information recovery from the BAO feature have involved `reconstruction' schemes that attempt to (partially) undo the nonlinear processing of the BAO feature, thereby sharpening it and bringing it closer to the pristine prediction of linear theory \cite{eisenstein+07}. Reconstruction schemes have either relied on perturbative approximations \cite{me99,pwc09,padmanabhan+12,tz12,white15,bph15,schmittfull+15,he18} or leveraged advances in stochastic sampling and machine learning \cite{bst24,bst25,bayer+26} to describe and undo the effects of nonlinear evolution. In either case, the eventual extraction of cosmological information from the reconstructed BAO feature typically relies on modelling the shape of the feature using templates derived from a chosen cosmological model such as $\Lambda$ cold dark matter ($\Lambda$CDM) \cite{blake+11,kirkby+13,cuesta+16,beutler+17,chen+24,DESI-DR2-II-BAOcosmo}.

An alternative line of research has explored a more model-agnostic extraction of cosmological information \cite{nsz21a,nsz21b,shapefit2021,ps23,fullShapeXir,novell-masot+25-DESIPkBk}. Here, one typically assumes a generic model of nonlinear evolution such as the Zel'dovich approximation \cite{zeldovich70}, which is accurate at the large scales relevant for the BAO feature. The Zel'dovich approximation predicts that gravitationally driven bulk flows, which are coherent at comoving scales of $\lesssim100\Mpch$ (coincidentally similar to the BAO scale) will tend to smear out the BAO feature by an approximately Gaussian kernel of width $\lesssim10\Mpch$ \cite{bharadwaj96,cs06a,cs08}. Additional ingredients are typically necessary, in the form of model-agnostic descriptions of the shape of the \emph{linear theory} dark matter 2pcf \cite{nsz21a,ps22,ps25a,fullShapeXir} and the impact of nonlinear evolution on the clustering of biased tracers \cite{ps25b}.

Starting in \cite{ps23} (hereafter, PS23), we have begun developing such a model-agnostic inference framework for the redshift-space BAO feature, which we call the `Zel'dovich smearing approximation'. In \cite{ps25a} (hereafter, PS25a), we used machine learning techniques to develop an optimal \biseq\ basis set for describing the linear theory dark matter 2pcf. In \cite{ps25b} (hereafter, PS25b), we demonstrated the importance of modelling the effects of scale-dependent density and velocity bias, as well as mode coupling, at BAO scales, providing a minimal agnostic model motivated by the Zel'dovich approximation and peaks theory. In \cite{ps26a} (hereafter, PS26), we put together all these ingredients and showed using toy data that the resulting Zel'dovich smearing approximation can be self-consistently used for cosmological inference using redshift-space multipoles of the nonlinearly evolved BAO feature.

A final ingredient, which is common to all the BAO inference techniques mentioned above but was missing in our Zel'dovich smearing approximation, is the impact of the fiducial cosmological model that is inevitably needed to convert observed angles and redshifts in the sky into comoving positions and separations that can be compared with theoretical predictions. In this work, we complete this step and show how to account for the resulting geometrical distortions in the Zel'dovich smearing approximation. With this complete model in hand, we also improve upon the performance tests of \citetalias{ps26a} by conducting more stringent model-agnostic cosmological inference exercises on semi-realistic tracer samples drawn from the \abacussummit\ simulation suite \cite{maksimova+21-AbacusSummit}.

The paper is organized as follows. In section~\ref{sec:fiducial-impact}, we set up our notation and describe how the geometric distortions due to the choice of fiducial cosmology affect the observables relevant for the Zel'dovich smearing approximation. We also discuss how our approach differs from the traditional, template-based analyses due to its different treatment of the physics governing the sound horizon. Section~\ref{sec:sims} provides technical details of our analysis; this includes the fiducial cosmology (used for geometric conversions) and baseline cosmology (used for testing the framework), the tracer samples constructed from the \abacussummit\ suite, a description of the pairwise correlation measurements in configuration and Fourier space used as our observables, and details of our cosmological inference setup. Section~\ref{sec:results} discusses the results of our inference exercises, including a discussion of the constraining ability of our model-agnostic approach and a brief comparison with the traditional approach. We end in section~\ref{sec:conclude} with a discussion of how cosmological information is organized in the Zel'dovich smearing approximation and provide a roadmap to a detailed comparison with traditional, template-based inference. The Appendices present additional details of our analysis; Appendix~\ref{app:model} spells out the calculations underlying the model predictions reported in section~\ref{sec:fiducial-impact}, Appendix~\ref{app:priors} describes the priors used in our inference exercises, Appendix~\ref{app:covariance} describes the error covariance used to define our data likelihood and Appendix~\ref{app:allparams} shows detailed posterior distributions. 

\section{Model}
\label{sec:fiducial-impact}

\subsection{Notation}
\label{subsec:notation}
Let the true comoving pair separation be $\ss_{\rm t}=s_{\parallel{\rm t}}\hat n + s_{\perp{\rm t}}\hat p$, where $\hat n$ is a unit vector along the line of sight and $\hat p$ is a unit vector in the sky plane, so that $\hat n\cdot\hat n=1=\hat p\cdot\hat p$ and $\hat n\cdot\hat p=0$.\footnote{We assume the flat sky approximation throughout.} We then have
\beq
s_{\parallel{\rm t}}=cH^{-1}(z)\Delta z\,;\quad s_{\perp{\rm t}} = D_{\rm M}(z)\Delta\theta\,,
\eeq
where $z$ is the median redshift of the sample, $H(z)$ is the Hubble parameter and $D_{\rm M}(z)=(1+z)D_{\rm A}(z)$ is the comoving angular diameter distance to redshift $z$ in the ground truth cosmology, and $\Delta z$ and $\Delta\theta$ are the observed pair separations in redshift and angle, respectively.

When using a fiducial cosmology close to, but different than the ground truth, we interpret pair separations as 
\beq
\ss_{\rm f}=s_{\parallel{\rm f}}\hat n+ s_{\perp{\rm f}}\hat p = \alpha_\parallel s_{\parallel{\rm t}}\hat n + \aperp s_{\perp{\rm t}}\hat p \,,
\label{eq:pairsep-true-to-fid}
\eeq
where
\beq
 \alpha_\parallel(z) \equiv H_{\rm f}^{-1}(z)/H^{-1}(z)\,;\quad 
 \aperp(z) \equiv D_{\rm Mf}(z)/D_{\rm M}(z)\,,
\label{eq:alpha-def}
\eeq
where quantities with the subscript `f' are evaluated in the fiducial cosmology. For brevity, we will henceforth drop the dependence of the $\alpha$ variables on the survey redshift $z$.  

It is useful to define the `Alcock-Paczynski' parameter
\beq
\aAP \equiv \aperp / \alpha_\parallel\,,
\label{eq:alpha_AP-def}
\eeq
in terms of which we can write the relations between the true magnitude $s_{\rm t} \equiv\sqrt{s_{\parallel{\rm t}}^2+s_{\perp{\rm t}}^2}$ and angle cosine $\mu_{s{\rm t}}\equiv (\ss_{\rm t}\cdot\hat n)/s_{\rm t} = s_{\parallel{\rm t}}/s_{\rm t}$ and their correspondingly defined fiducial counterparts $s_{\rm f},\mu_{s{\rm f}}$ as (e.g., section 2.2 of \cite{perez-fernandez+25})
\begin{align}
s_{\rm t} &= \frac{s_{\rm f}}{\aperp} \sqrt{1 + \mu_{s{\rm f}}^2\left(\aAP^2-1\right)} \,, \label{eq:s(sf,musf)}\\
\mu_{s{\rm t}} &= \frac{\aAP\mu_{s{\rm f}}}{\sqrt{1 + \mu_{s{\rm f}}^2\left(\aAP^2-1\right)}} \,.
\label{eq:mus(sf,musf)}
\end{align}
Below, we will also make use of the isotropized scale parameter \aiso\ defined by
\beq
\aiso \equiv \left(\alpha_\parallel\aperp^2\right)^{1/3} = \aperp \aAP^{-1/3}\,,
\label{eq:alpha_iso-def}
\eeq
which can also be written as a ratio of isotropized comoving distances $D_{\rm Vf}$ and $D_{\rm V}$,
\beq
\aiso = D_{\rm Vf}(z)/D_{\rm V}(z)\,,
\label{eq:aiso-as-ratio}
\eeq
with
\beq
D_{\rm V}(z) \equiv \left(\frac{cz}{H(z)}\,D_{\rm M}(z)^2\right)^{1/3}\,; \quad D_{\rm Vf}(z) \equiv \left(\frac{cz}{H_{\rm f}(z)}\,D_{\rm Mf}(z)^2\right)^{1/3}\,.
\label{eq:DV-def}
\eeq
We will see below that the impact of the choice of fiducial cosmology can, in principle, be written in terms of two parameters \DaAP\ and \Daiso, where
\beq
\DaAP \equiv \aAP-1\,; \quad \Daiso\equiv \aiso-1\,.
\label{eq:Daiso,DaAP-def}
\eeq
In fact, we will argue that, in our model-agnostic framework, only \DaAP\ is relevant.
In the notation of \cite{xu+13} (see their section 2.1), we have $\aiso=\alpha^{-1}(r_{\rm sf}/r_{\rm s})$ (where $r_{\rm s}$ is the sound horizon) and $\aAP=(1+\epsilon)^3$, so that $\DaAP\simeq3\epsilon$.

The Zel'dovich smearing approximation of \citetalias{ps26a} models two kinds of pairwise correlation measurements near the BAO scale, the anisotropic configuration space 2pcf $\xi$ and low-$k$ integrals $\Sigma^2$ of the anisotropic power spectrum in Fourier space. In each case, we focus on estimates of the multipoles of these quantities, namely $\xiell{\ell}(s_{\rm t})$ for the 2pcf and $\Sigell{\ell}$ for the power spectrum integrals, for $\ell=0,2,4$ (respectively, the monopole, quadrupole and hexadecapole). In this work, we additionally include the impact of the choice of fiducial cosmology so as to model the final observables $\xiellobs{\ell}(s_{\rm f})$ and $\Sigellobs{\ell}$.

\subsection{Predictions for observables}
\label{subsec:prediction}
In Appendix~\ref{subapp:2pcf}, we show that the relation between $\xiellobs{\ell}(s_{\rm f})$ and $\xiell{\ell}(s_{\rm t})$ can be approximated as
\begin{align}
\xiellobs{\ell}(s_{\rm f}) &\simeq \xiell{\ell}(s_{\rm f}) + \DaAP \bigg[\frac23\left(2\ell+1\right) \sum_{\ell^\prime}\Cal{C}_{\ell\ell^\prime}\,s_{\rm f}\p_{s_{\rm f}}\xiell{\ell^\prime}(s_{\rm f}) + \sum_{\ell^\prime}\Cal{A}_{\ell\ell^\prime}\, \xiell{\ell^\prime}(s_{\rm f}) \bigg] \,,
\label{eq:xiobs(ell)}
\end{align}
where the r.h.s. involves the undistorted $\xi^{(\ell)}$ and their derivatives, evaluated at the distorted $s_{\rm f}$, with the matrices $\Cal{C}_{\ell\ell^\prime}$ and $\Cal{A}_{\ell\ell^\prime}$ being respectively defined in \eqns{eq:Cmatrix} and \eqref{eq:Amatrix}. Moreover, the r.h.s. involves \DaAP\ but not \Daiso, for the reasons discussed in section~\ref{subsec:compare-traditional} and Appendix~\ref{subapp:2pcf}.
 
Similarly, Appendix~\ref{subapp:Pk} shows that \Sigellobs{\ell} and \Sigell{\ell} can be related as
\begin{align}
\Sigellobs{\ell} &\simeq \Sigell{\ell} + \DaAP \sum_{\ell^\prime} \left(\Cal{A}_{\ell\ell^\prime} - \frac43(2\ell+1)\,\Cal{C}_{\ell\ell^\prime}\right)  \Sigell{\ell^{\prime}} \,,
\label{eq:Sig2ell-obs}
\end{align}
which \emph{also} does not depend on \Daiso, provided we assume that the model actually predicts $\Sigell{\ell}\times\aiso^2$. Importantly, we see that all the terms in \eqns{eq:xiobs(ell)} and~\eqref{eq:Sig2ell-obs} are either already predicted or easily derived in the Zel'dovich smearing approximation. 

\subsection{Comparison with traditional BAO analysis}
\label{subsec:compare-traditional}
The quantity \DaAP\ (equation~\ref{eq:Daiso,DaAP-def}) appears as a new, cosmologically relevant parameter in \eqns{eq:xiobs(ell)} and~\eqref{eq:Sig2ell-obs}. While this is similar to what happens in a traditional, template-based BAO analysis, there are important differences which we discuss next.

One major difference with a traditional analysis is the absence of \Daiso\ in \eqns{eq:xiobs(ell)} and~\eqref{eq:Sig2ell-obs}. 
In fact, a calculation of $\xiellobs{\ell}(s_{\rm f})$ from first principles \emph{does} lead to such a dependence, as we show in Appendix~\ref{subapp:2pcf} where \eqn{eq:xiobs(ell)-sbins} contains a term proportional to \Daiso.
This term captures the leading order scaling behaviour of pair separations, which is a proportionality to $s/D_{\rm V}$. The primordial location of any feature in the 2pcf, e.g., $r_{\rm LP}$, then also inherits this scaling \cite{anselmi+19}. In traditional BAO analyses, this is accounted for by including the BAO sound horizon $r_{\rm s}$ in the definition of \aiso\ (e.g., see the comment below equation~\ref{eq:Daiso,DaAP-def}). 

In the Zel'dovich smearing model, however, all primordial scales $r$ are described by the basis coefficients $\{w_m\}$ of the \biseq\ basis used to model the linear 2pcf. These are \emph{a priori} independent of \Daiso, since our definition of \aiso\ \emph{does not} involve the sound horizon (see equation~\ref{eq:aiso-as-ratio}). Any scaling introduced by \Daiso\ could then be precisely compensated for in the model by suitably adjusting the values of $\{w_m\}$, leading to unbroken degeneracies between these parameters.
We have indeed checked that the resulting strong degeneracies substantially increase the error on derived parameters such as $r_{\rm LP}$, unless we invoke undesirably tight priors.\footnote{This would also be the case in a traditional analysis, were we to try and separately constrain $D_{\rm V}$ and the sound horizon $r_{\rm s}$ (e.g., Table~I of \cite{anselmi+19}).}

Unlike a traditional, template-based analysis, where the sound horizon $r_{\rm s}$ is a derived parameter, our model-agnostic framework does not explicitly involve $r_{\rm s}$, so we cannot directly account for its degeneracy with $D_{\rm V}$. 
To overcome this problem, we instead allow the coefficients $\{w_m\}$ describing the primordial physics to absorb the geometrical effect of the $D_{\rm V}$ scaling by \emph{declaring} the 2pcf to be functions of $y=s/D_{\rm V}$ instead of $s$ (see, e.g., \cite{anselmi+19} and Appendix~\ref{subapp:2pcf}) and similarly for the power spectrum. This makes our analysis dual, in a sense, to the traditional one; there, it is the geometrical factor \aiso\ that absorbs the physics of the sound horizon by defining it as a ratio of $D_V/r_s$ values in the true and fiducial cosmologies, rather than only a ratio of $D_V$ values as we do.

For consistency, when reporting the results for any primordial scale $r$ in our parameter inference exercises, we will display the posterior distribution of $r$ as predicted by the model and compare it to the ground truth value of $r_{\rm true}\times D_{\rm Vf}/D_{\rm V,true}=r_{\rm true}(1+\Delta\alpha_{\rm iso,true})$. This includes not only the length scales derived from the the basis coefficients $\{w_m\}$, but also scales like \sigv\ which predict \Sigellobs{2}. This is consistent with the discussion at the end of section~\ref{subsec:prediction}.

\section{Simulations and analysis}
\label{sec:sims}
In this work, we use simulated halo positions from the publicly available \abacussummit\footnote{\url{https://abacussummit.readthedocs.io/en/latest/abacussummit.html}} suite of simulations \cite{maksimova+21-AbacusSummit} to produce semi-realistic tracer samples mimicking those expected to be observed by the DESI and \emph{Euclid} surveys. In this section, we describe these samples and the resulting measurements, along with details of the inference technique including the definition of priors, estimates of the covariance matrix for the likelihood and the sampling methodology. 

\subsection{Fiducial and baseline cosmologies}
\label{subsec:cosmologies}
We focus on tracer samples in the baseline \base\ cosmology of the \abacussummit\ suite, which is a flat $\Lambda$CDM model with one massive and two massless neutrino species, consistent with Planck 2018 results \cite{Planck18-VI-cosmoparam}. The cosmological parameters are $\Omega_{\rm m} = 0.3138$, $\Omega_{\rm b} = 0.04930$, $h = 0.6736$, $n_{\rm s} = 0.9649$, $\ln(10^{10}A_{\rm s}) = 3.0364$, $N_{\rm ur}=2.038$, $m_\nu=0.06\,$eV. This functions as the ground truth for our analysis.

Our fiducial cosmology is also a flat $\Lambda$CDM model consistent with Planck 2018, but with three massless neutrino species. The cosmological parameters are $\Omega_{\rm m} = 0.3153$, $\Omega_{\rm b} = 0.04929$, $h = 0.6737$, $n_{\rm s} = 0.9649$, $\ln(10^{10}A_{\rm s}) = 3.0450$, $N_{\rm ur}=3.044$. This was the cosmology used by \citetalias{ps25a} for calibrating their \biseq\ basis, and functions as the cosmology using which angles and redshifts are converted to distances in our analysis.

Wherever needed, we generate cosmological transfer functions for matter fluctuations using the \textsc{class} code \cite{class-I,class-II}.\footnote{\url{http://class-code.net/}}

\subsection{Simulation configuration and halo catalogs}
\label{subsec:catalogs}
The \abacussummit\ suite provides 25 realizations of its baseline \base\ cosmology whose parameters were described above. Each box is periodic, with volume $(2\Gpch)^3$, and was simulated with $6192^3$ particles. The resulting particle mass of $\sim2\times10^9\Mh$ means that LRG-like halos $\gtrsim10^{12.8}\Mh$ are resolved with $\gtrsim {\rm few} \times 10^3$ particles. Halos were identified using the spherical overdensity \textsc{CompaSO} algorithm\footnote{\url{https://abacussummit.readthedocs.io/en/latest/compaso.html}} \cite{hadzhiyska+22-Compaso}.

We consider two samples constructed using a single \abacussummit\ box at a time, so as to mimic (a) at redshift $z=0.8$, the LRG population being observed by the DESI survey (we refer to this as the \desi\ sample below) and (b) at redshift $z=1.1$, the ELG population being observed by the \emph{Euclid} mission (hereafter, \euclid). Our effective comoving volume is then $V_{\rm eff} \simeq 26.2\, {\rm Gpc}^3$, somewhat larger than the expected value for the final observed DESI LRG-2 sample and nearly the same as that expected for the observed \emph{Euclid} ELG sample in the redshift range $1.0\leq z\leq1.2$. To model the impact of cosmic variance between the \desi\ and \euclid\ samples, we use different realizations for generating samples for each of these, namely \texttt{ph000} for \desi\ and \texttt{ph009} for \euclid. In reality, there is expected to be significant overlap in sky coverage for the two surveys (e.g., \cite{naidoo+23}) which will be important for cross-correlation, but we ignore this here for simplicity.

\begin{table}
\centering
\begin{tabular}{ccccccc}
\hline\hline
Configuration & Redshift & $M_{\rm halo,min}$ & Volume & $N_{\rm halos}$ & \abacussummit\ & Ref. \\
 & $z$ & $(10^{12}\Mh)$ & $(\Gpch)^3$ & $(\times10^6)$ & phase &  \\
\hline\hline
\desi\ & $0.8$ & $8.0$ & $2.0$ & $2.59$ & \texttt{ph000} & Fig.~\ref{fig:pwise-desiLRG2} \\
\euclid\ & $1.1$ & $1.0$ & $2.0$ & $6.50^\dagger$ & \texttt{ph009} & Fig.~\ref{fig:pwise-euclidELG} \\
\hline\hline
\end{tabular}
\caption{Simulation configurations constructed from the \abacussummit\ suite and used in this work. Both configurations used the baseline \texttt{c000} cosmology. ${}^\dagger$The number reported for the \euclid\ sample is after applying a random downsampling by factor $4$. See text for details.}
\label{tab:sims}
\end{table}

We impose mass thresholds $M_{\rm halo}\geq8\times10^{12}\Mh$ for \desi\ and $M_{\rm halo}\geq 10^{12}\Mh$ for the \euclid\ sample. The mass threshold for \desi\ is consistent with values expected for LRGs at these epochs (e.g., see Table~2 of \cite{zhai+17}), while that for \euclid\ is approximately consistent with the \emph{Euclid} Flagship mock galaxy catalog \cite{euclid-flagship25} (see their section~6.2) and early results from DESI \cite{favole+25}. Due to computational limitations, we randomly downsample our \euclid\ sample by a factor 4. This does not affect the quality of our measurements since these are at large enough length scales to be dominated by cosmic variance. Our \desi\ and \euclid\ samples then respectively contain $\sim2.6$ million and $\sim6.5$ million halos. These numbers are summarized in Table~\ref{tab:sims}. For comparison, the DESI DR2 LRG-2 sample at $z\sim0.7$ contains $\sim1.6$ million objects in an effective comoving volume $\sim7.6\,{\rm Gpc}^3$ (see Table~III of \cite{DESI-DR2-II-BAOcosmo}).

\subsection{Pairwise correlation measurements}
\label{subsec:measurements}
For the redshift space observables required in this work, the halo positions in each simulation box are processed as follows, using the notation set up in section~\ref{subsec:notation}. First, assuming one of the coordinate axes to be aligned with the line of sight, the halo positions are modified to include the effects of redshift space distortions, 
\beq
s_{\parallel{\rm t}} \to s_{\parallel{\rm t}} + \frac{v_\parallel(1+z)}{H(z)}\,,
\label{eq:box-rsd}
\eeq
where $s_{\parallel{\rm t}}$ is a comoving position in the true cosmology, $v_\parallel$ is the tracer's proper velocity component in \kms\ along the line of sight and $H(z)$ is the ground truth Hubble parameter at the sample redshift. Additionally, the halo positions are also converted to positions in the fiducial cosmology using \eqn{eq:pairsep-true-to-fid}. Under the flat sky approximation that we assume, this is equivalent to converting the angles and redshifts produced by the ground truth into comoving separations using the fiducial cosmology \cite{pw08,chen+24,perez-fernandez+25}. The conversion from the \abacussummit\ baseline \base\ cosmology to our fiducial Planck 2018 cosmology for the samples at $z=0.8$ and $z=1.1$ gives us
\begin{align}
\alpha_\parallel-1 &= 1.089 \times 10^{-3}\, (z=0.8)\,,\,\, 1.329\times 10^{-3}\, (z=1.1)\notag\\
\alpha_\perp-1 &= 4.939 \times 10^{-4}\, (z=0.8)\,,\,\, 6.458\times 10^{-4}\, (z=1.1)\,.
\end{align}
The corresponding ground truth values of \DaAP\ (equations~\ref{eq:alpha_AP-def} and \ref{eq:Daiso,DaAP-def}) and \Daiso\ (equations~\ref{eq:alpha_iso-def} and \ref{eq:Daiso,DaAP-def}) are $\sim10^{-4}\,(10^{-3})$ at $z=0.8\,(1.1)$.\footnote{Note that, although \Daiso\ will eventually not appear in our analysis for the reasons discussed in section~\ref{subsec:compare-traditional}, its effect is implicitly included when generating our mock observations; this closely mimics what would happen in actual observations.}  
All positions are finally converted to units of \Mpch\ in the fiducial cosmology.

\emph{For brevity, hereon we will drop the subscript `f' on separations and wavenumbers in the fiducial cosmology.} 

The anisotropic 2pcf $\hat\xi_{\rm obs}(s,\mu_s)$ is estimated using the redshift-space distorted positions along with the geometrical conversion mentioned above, wrapping periodically in the cubic box and applying the Peebles-Hauser estimator in 2 dimensions:
\beq
\hat\xi_{\rm obs}(s,\mu_s) = \frac{DD(s,\mu_s)}{RR(s,\mu_s)} - 1\,,
\eeq
where $DD(s,\mu_s)$ is the normalised count of tracer pairs in the separation interval $(s,s+\Delta s)$ and $(\mu_s,\mu_s+\Delta\mu_s)$, and $RR(s,\mu_s)$ is the corresponding count for random pairs, which takes the analytical form
\beq
RR(s,\mu_s) = \Delta\mu_s\,\frac{2\pi}{3L^3}\left((s+\Delta s)^3-s^3\right)
\eeq
for a periodic box of length $L$. 

To estimate the corresponding anisotropic power spectrum $\hat P_{\rm obs}(k,\mu_k)$, we first estimate a tracer number density field using cloud-in-cell (CIC) interpolation on the redshift-space distorted, geometrically modified tracer positions on a $256^3$ grid.\footnote{We have checked that, for the large scale (low-$k$) measurements needed in this work, our results are fully converged with respect to grid size.} Fourier transforming the CIC field leads to the $k$-space overdensity field $\hat\delta(\mathbf{k})$, using which the power spectrum is estimated by averaging $|\hat\delta(\mathbf{k})|^2$ in bins of $k=|\mathbf{k}|$ and $\mu_k=(\mathbf{k}\cdot\hat n)/k$.

In each case, the required multipoles are then estimated by numerically integrating the 2-d measurements over the respective cosine angle, weighted by Legendre polynomials:
\begin{align}
\xiellsim{\ell}(s) &= \frac{(2\ell+1)}{2} \,\Delta\mu \sum_j \hat\xi(s,\mu_j)\,\Cal{P}_\ell(\mu_j)\,, 
\label{eq:2pcf-estimate}\\
\hat P_{\rm obs}^{(\ell)}(k) &= \frac{(2\ell+1)}{2}\,\Delta\mu_k\sum_j\hat P(k,\mu_{kj})\,\Cal{P}_\ell(\mu_{kj})\,,
\label{eq:Pk-estimate}
\end{align}
where the sum is over bins of $\mu$ or $\mu_k$, and we use 161 linearly spaced bins in the range $[-1,1]$ in each case.
\begin{figure}
\centering
\includegraphics[width=0.49\linewidth]{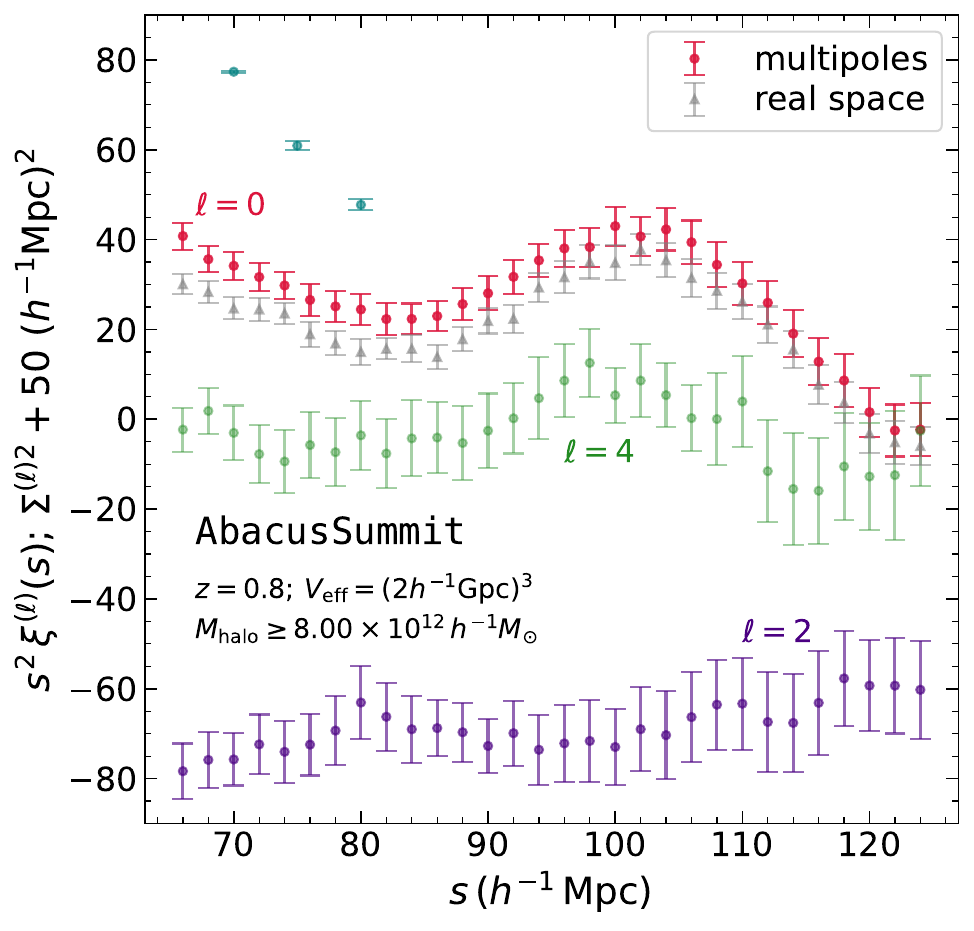}
\includegraphics[width=0.49\linewidth]{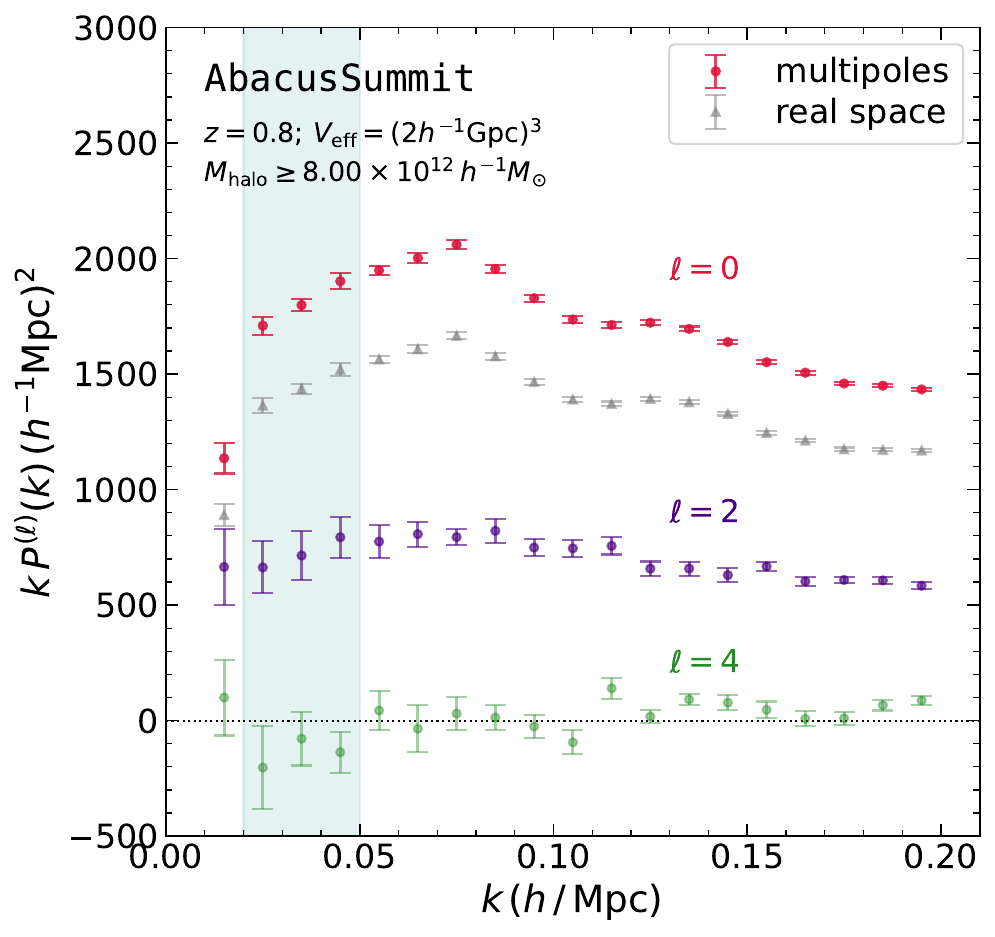}
\caption{{\bf Pairwise correlation measurements for the \desi\ sample constructed using \abacussummit\ halos.} \emph{(Left panel):} 2pcf multipoles $\xiellsim{\ell}(s)$ (colour-coded circles with error bars), power spectrum multipole integrals $\Sigellsim{\ell}$ (teal circles with error bars) and real space 2pcf (gray triangles with error bars). For clarity, we have given an additive offset of $+50(\Mpch)^2$ to each of the \Sigellsim{\ell} values.
\emph{(Right panel):} Power spectrum multipoles $\hat P^{(\ell)}(k)$ (colour-coded circles with error bars) and real space power spectrum (gray triangles with error bars). 
Vertical band encloses the bins used for the \Sigellsim{\ell} measurements shown in the \emph{left panel}. 
Text labels give details of the sample definition, see also Table~\ref{tab:sims}.
All length scales are in units of \Mpch\ where $h=h_{\rm fid}=0.6737$. 
}
\label{fig:pwise-desiLRG2}
\end{figure}

Following \citetalias{ps26a}, we use $30$ linearly spaced bins in $s$ of width $2\Mpch$ for separations in the range $65\leq s/(\Mpch) \leq 125$\footnote{The upper limit of $125\,\Mpch$ is slightly larger than the value $120\,\Mpch$ used by \citetalias{ps25a} to calibrate the \biseq\ basis. We have checked that this difference does not affect any of our results. Our choice also matches that used in the toy models of \citetalias{ps26a}.} in \eqn{eq:2pcf-estimate}, after which the 2pcf multipoles $\xiellsim{\ell}(s)$ are ready for use. The integrals \Sigellsim{\ell} over the power spectrum multipoles are estimated as
\beq
\Sigellsim{\ell} = \frac{\Delta k}{6\pi^2}\sum_{k_{\rm min}\leq k_i\leq k_{\rm max}}\,\hat P^{(\ell)}_{\rm obs}(k_i)\,,
\label{eq:Sig2ell-estimate}
\eeq
where the sum is over 3 linearly spaced bins in $k$ with $\Delta k=0.01\hMpc$, $k_{\rm min}=0.02\hMpc$ and $k_{\rm max}=0.05\hMpc$ as motivated by \citetalias{ps26a}.

\begin{figure}
\centering
\includegraphics[width=0.49\linewidth]{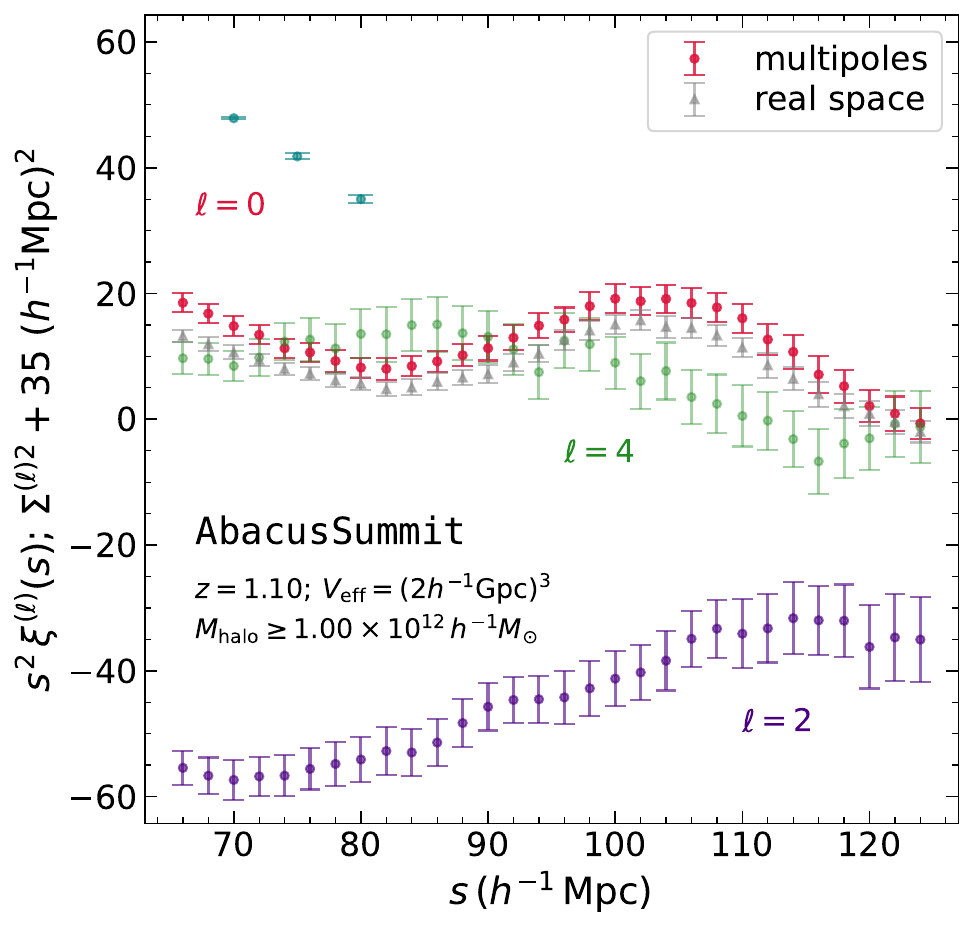}
\includegraphics[width=0.49\linewidth]{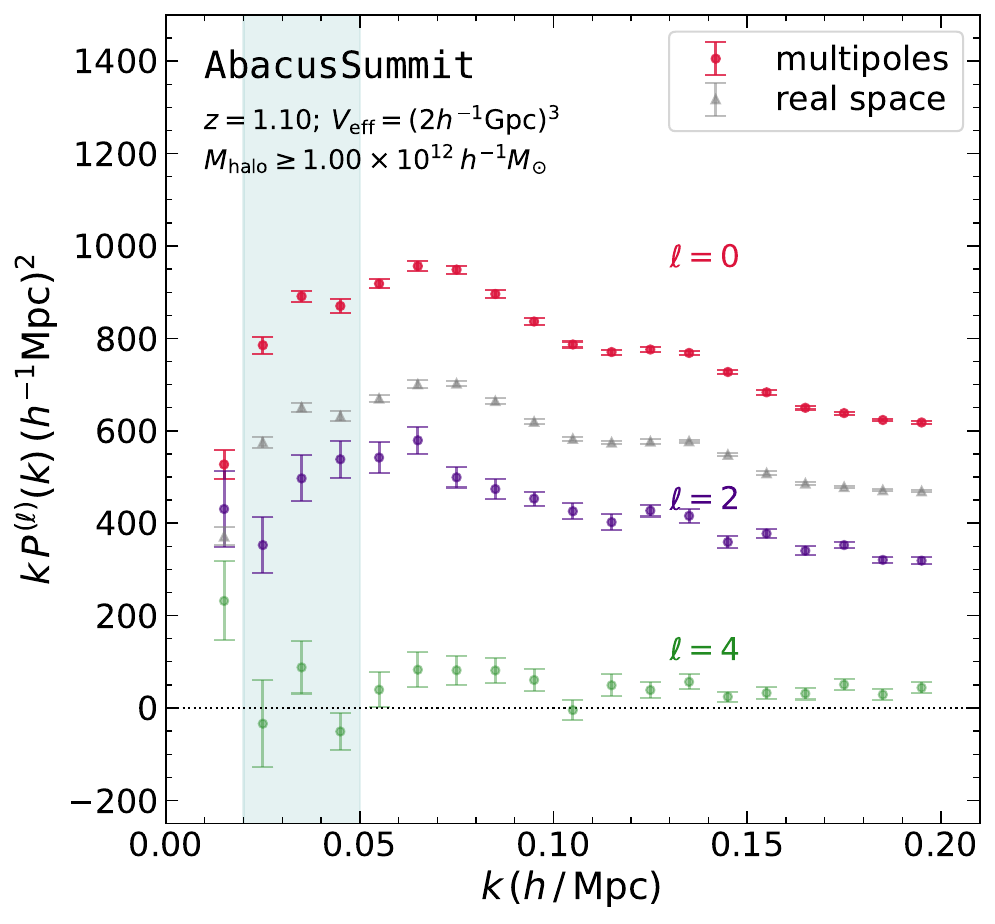}
\caption{Same as Fig.~\ref{fig:pwise-desiLRG2}, showing {\bf measurements for the \euclid\ sample.} Note the $y$-axis scales which differ from those in Fig.~\ref{fig:pwise-desiLRG2}, primarily due to the reduced value of the linear Eulerian bias $b$. For the same reason, the additive offset to the \Sigellsim{\ell} values in the \emph{left panel} is now reduced to $+35(\Mpch)^2$. 
}
\label{fig:pwise-euclidELG}
\end{figure}

Figs.~\ref{fig:pwise-desiLRG2} and~\ref{fig:pwise-euclidELG} display these pairwise measurements, using the \desi\ and \euclid\ samples, respectively. The \euclid\ measurements can be compared with figs.~41-43 of \cite{euclid-flagship25}. The \emph{left panels} of the Figures show the 2pcf multipoles $\xiellsim{\ell}(s_{\rm f})$ and the power spectrum multipole integrals \Sigellsim{\ell}, which are used later in the MCMC analysis. For completeness, we also display the real space 2pcf in the \emph{left panels}, while the \emph{right panels} show the power spectrum (real space and multipoles) for the same samples. These are estimated similarly to their anisotropic counterparts, but without moving to redshift space as in \eqn{eq:box-rsd} and without binning in $\mu_s$ or $\mu_k$.

For the cosmological inference exercise below, we modify the configuration space multipoles as discussed by  \citetalias{ps26a} and given in their equations (2.13)-(2.15), which we reproduce here for reference,
\begin{align}
\Delta \xiellsim{0}(s) &\equiv \xiellsim{0}(s) \,, 
\label{eq:DxiNL(0)-def} \\
\Delta \xiellsim{2}(s) &\equiv \xiellsim{2}(s) - (s_{\rm min}/s)^3\,\xiellsim{2}(s_{\rm min}) \,, \label{eq:DxiNL(2)-def} \\
\Delta \xiellsim{4}(s) &\equiv \xiellsim{4}(s) - (s_{\rm min}/s)^5\,\xiellsim{4}(s_{\rm min}) \,. \label{eq:DxiNL(4)-def} 
\end{align}
As discussed in Appendix~\ref{app:covariance}, we also make corresponding modifications in the error covariance matrix.

\subsection{Inference}
\label{subsec:inference}
Our parameter set has a total of 
19 parameters given by the set
\beq
\left\{\beta,\sigv,\{w_m\}, \DaAP, f_{\rm v},b,B_{1\ast},B_{v\ast},\sigma,A_{\rm MC},\qbar{2}{}\right\}\,.
\label{eq:fullparamset}
\eeq
Of these, \DaAP\ was introduced in section~\ref{sec:fiducial-impact} and the remaining 18 are the same as described by \citetalias{ps26a}. Briefly, $\beta=f/b$ is the usual cosmological RSD parameter, where $f=\der\ln D/\der\ln a$ is the linear growth rate with growth factor $D(a)$, \sigv\ is the linear theory 1-particle velocity dispersion, $\{w_m\}$ for $0\leq m\leq 8$ are the coefficients of the \biseq\ basis for describing the real space linear 2pcf $\xi_{\rm lin}(r)$ and $f_{\rm v}$ is the fraction of $\sigv^2$ included in the Fourier-space integration range used for estimating \Sigell{\ell}. Together, the 13 parameters $\left\{\beta,\sigv,\{w_m\},\DaAP,f_{\rm v}\right\}$ account for all the cosmological information described in the Zel'dovich smearing approximation. Of the remaining parameters, $b$ is the scale independent linear Eulerian bias of the tracer sample, while the subset $\left\{B_{1\ast},B_{v\ast},\sigma,A_{\rm MC}\right\}$ describes the effects of scale dependent bias and mode coupling, or \emph{sdbmc}. Although these are not cosmologically relevant, they are essential to include in the analysis since they display important degeneracies with the cosmological parameters. Finally, \qbar{2}{} is a nuisance parameter needed when modelling the hexadecapole \xiellobs{4} that arises from the necessarily finite range of separation scales modelled. We kindly refer the reader to \citetalias{ps26a} for the parameter definitions.

We perform parameter inference using the Monte Carlo Markov Chain (MCMC) technique, assuming a Gaussian likelihood throughout, with priors set as described in Appendix~\ref{app:priors} and a data covariance as described in Appendix~\ref{app:covariance}. Below, we use the publicly available \textsc{Cobaya} \citep{tl19-cobaya,tl21-cobaya}\footnote{\url{https://cobaya.readthedocs.io/}} and \textsc{GetDist} \citep{lewis19}\footnote{\url{https://getdist.readthedocs.io/}} packages to implement the MCMC and display results, respectively, discarding the first $30\%$ of the samples as burn-in. We run 6 chains in parallel and test for convergence using the generalized version of the Gelman-Rubin statistic described by \cite{lewis13} (this is automatically used by the \texttt{mcmc} sampler in \textsc{Cobaya}), demanding $R-1\leq0.01$ for the means and $R-1\leq0.04$ for the $95\%$ confidence regions.

\begin{figure}
\centering
\includegraphics[width=0.49\textwidth]{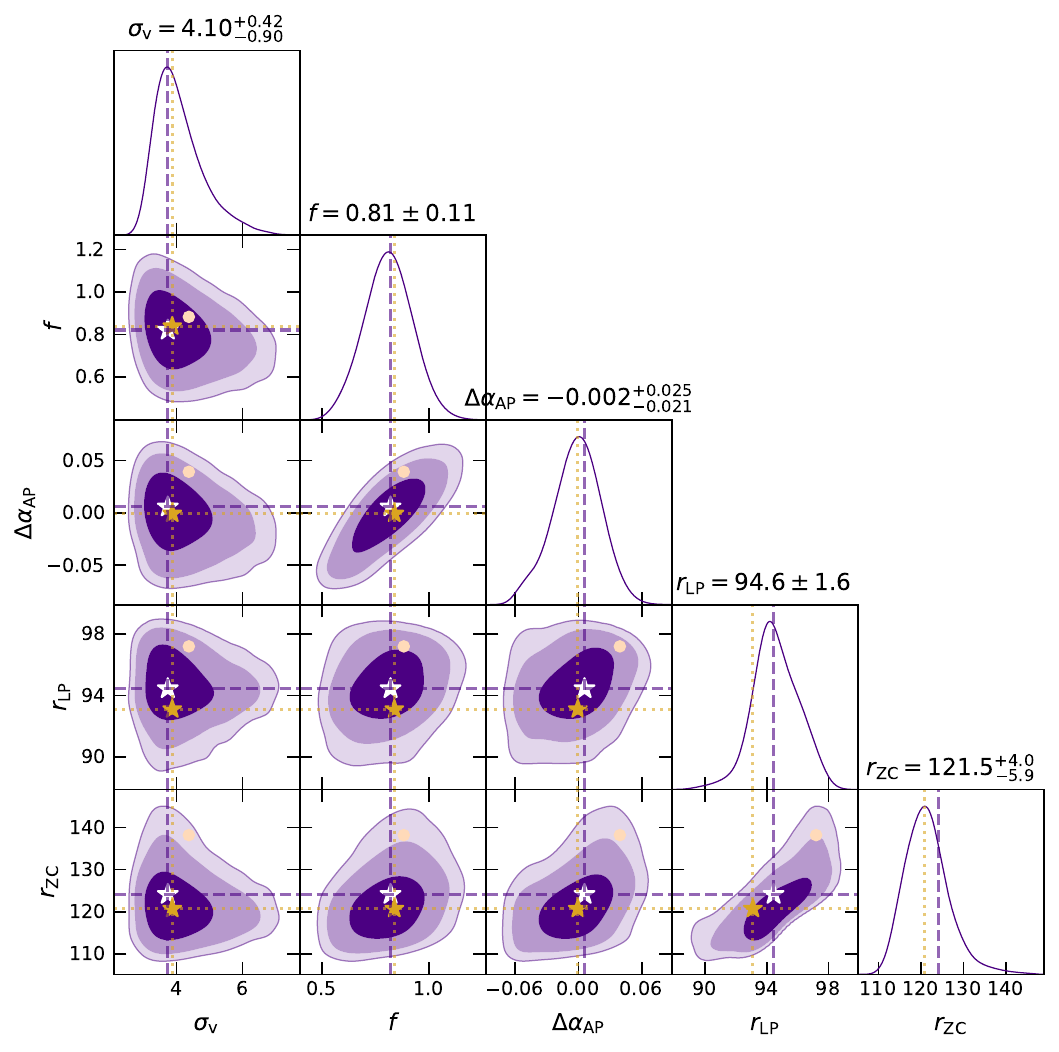}
\includegraphics[width=0.49\textwidth]{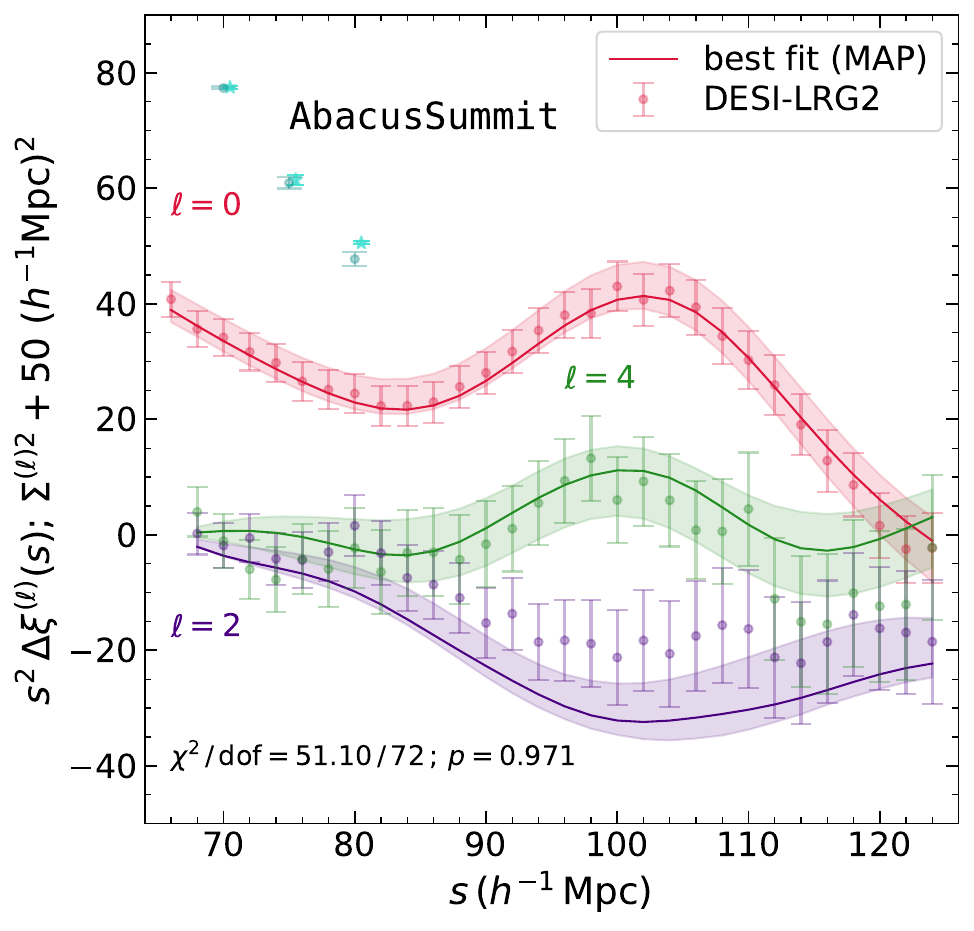}
\caption{{\bf Inference results for the \desi\ sample.} \emph{(Left panel):} Constraints on cosmological parameters. Contours show the $68\%,95\%,99\%$ confidence regions. Dashed lines intersecting at white stars indicate the best fit (i.e., maximum \emph{a posteriori}; MAP) parameter vector. Dotted lines intersecting at yellow stars indicate ground truth values. Peach circles indicate the maximum likelihood parameter vector. The diagonal panel titles give the median and $68\%$ confidence interval of the marginal constraints. We see excellent, unbiased recovery of all parameters at better than $68\%$ confidence. See Fig.~\ref{fig:contours-desiLRG2} for joint constraints on all varied parameters and Fig.~\ref{fig:contours-cosmoprior-desiLRG2} for a comparison of the posterior and weak $\Lambda$CDM prior on selected cosmological parameters. \emph{(Right panel):} Comparison of the data with the best fit model. Points with error bars show the mock data for $\Delta\xiellsim{\ell}(s)$ (colour-coded for $\ell$ as indicated by the text labels) and \Sigellsim{\ell} (three blue points in the upper left, with $\ell=0,2,4$ from left to right). These are derived from the measurements shown in the \emph{left panel} of Fig.~\ref{fig:pwise-desiLRG2} using \eqns{eq:DxiNL(0)-def}-\eqref{eq:DxiNL(4)-def}. Colour-coded solid curves with error bands show the best fit (MAP) and central $68\%$ confidence region from the parameter inference exercise for the 2pcf multipoles. The corresponding results for the power spectrum multipole integrals are shown by the cyan stars with asymmetric error bars. For clarity, we have given an additive offset of $+50 (\Mpch)^2$ to each of the $\hat\Sigma^{(\ell)2}$ values and their corresponding best fit results. 
The text label gives the value of $\chi^2$ for the maximum likelihood vector along with the number of degrees of freedom and corresponding $p$-value; the fit is of excellent quality (but with possibly overestimated errors, see discussion in text).}
\label{fig:cosmofit-desiLRG2}
\end{figure}


\section{Results}
\label{sec:results}
We organize our results similarly to those of \citetalias{ps26a}, for each of our tracer samples. In particular, in this section we compress the $\{w_m\}$ parameters into two interesting length scales, the linear point $r_{\rm LP}$ \cite{LP2016} and the zero-crossing $r_{\rm ZC}$ of the linear 2pcf $\xi_{\rm lin}(r)$. We show results for the full parameter set in Appendix~\ref{app:allparams}. 

\begin{figure}
\centering
\includegraphics[width=0.49\textwidth]{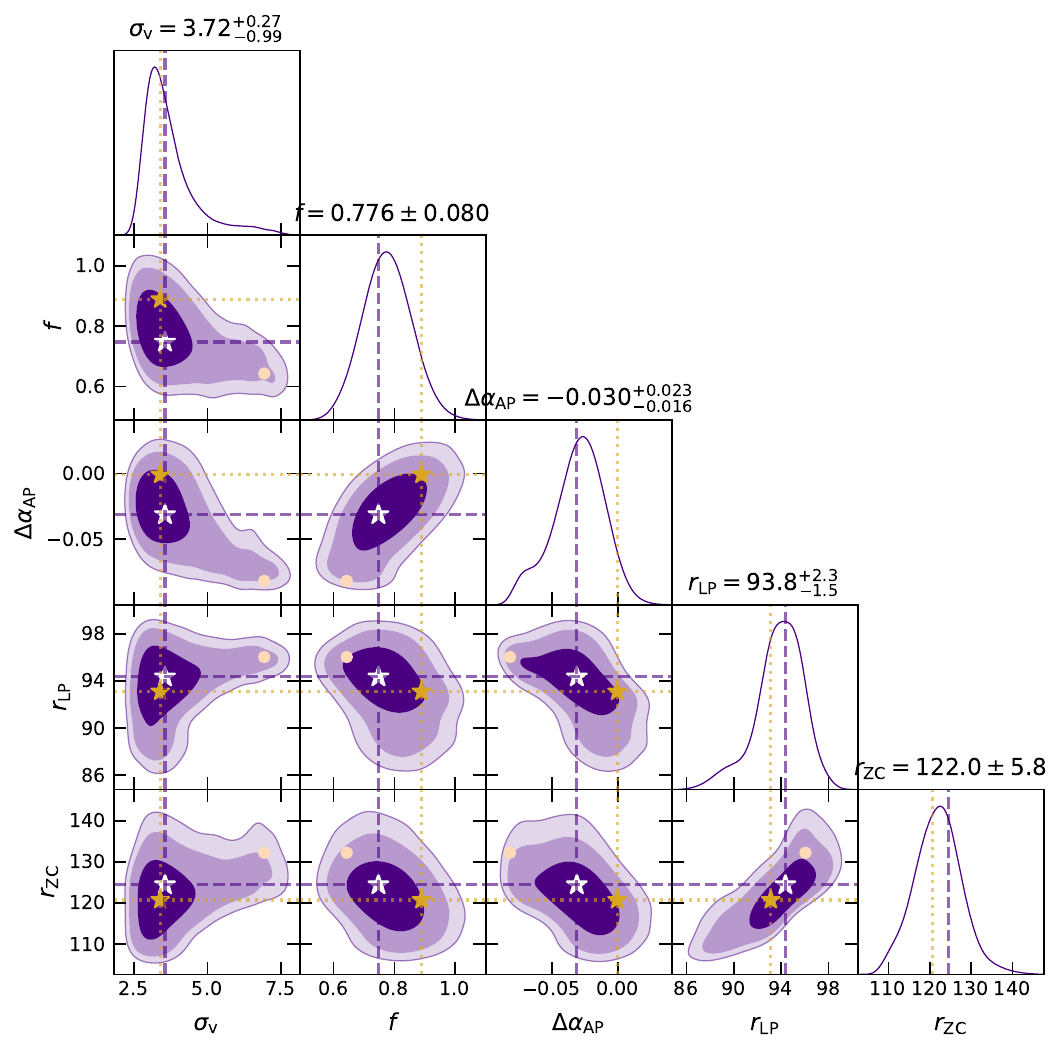}
\includegraphics[width=0.49\textwidth]{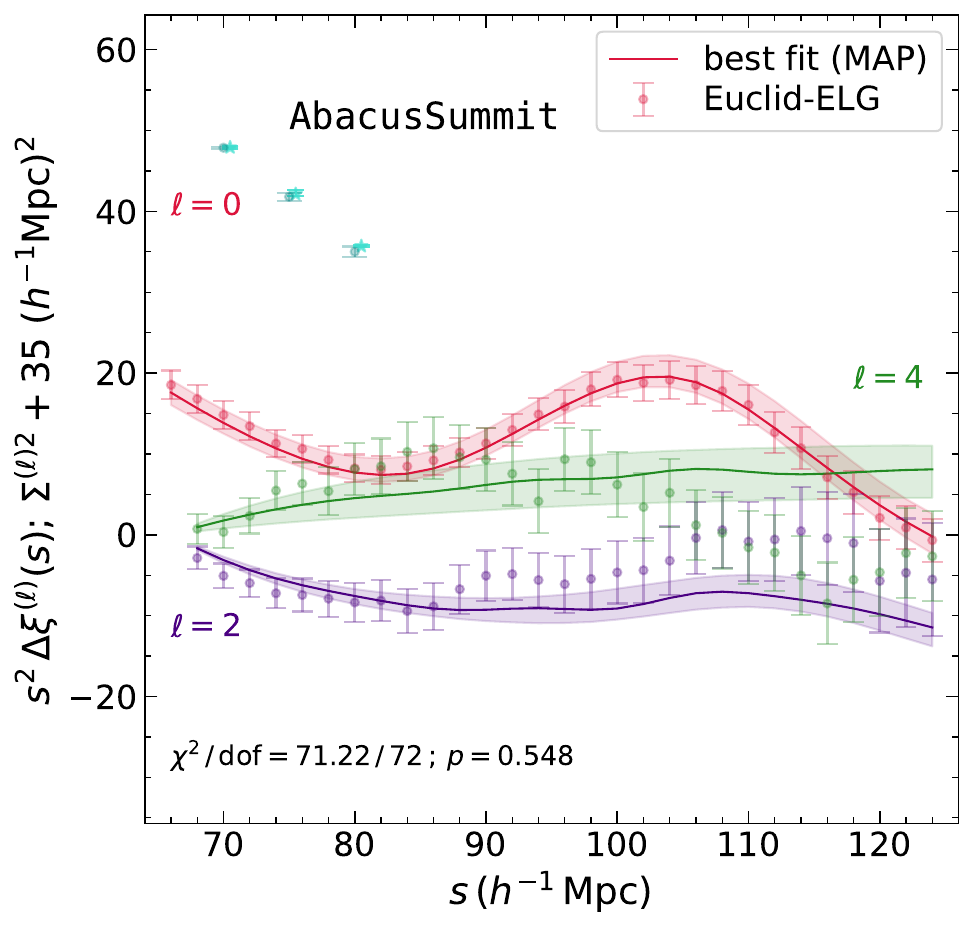}
\caption{Same as Fig.~\ref{fig:cosmofit-desiLRG2}, {\bf for the \euclid\ sample,} with data points derived from the measurements in the \emph{left panel} of Fig.~\ref{fig:pwise-euclidELG}.}
\label{fig:cosmofit-euclidELG}
\end{figure}


\subsection{Cosmological constraints}
\label{subsec:cosmo-constraints}
Fig.~\ref{fig:cosmofit-desiLRG2} shows the posterior for the cosmological parameters and compares the best fitting model with the data for the \desi\ sample. We see excellent, unbiased recovery of all parameters at better than $68\%$ confidence in the pairwise posterior distributions. The fit quality is also excellent, although the somewhat low value of $\chi^2/{\rm dof}\sim 0.71$ indicates that either the diagonal errors or the strength of the covariance (Appendix~\ref{app:covariance},  Fig.~\ref{fig:GPcov}), or both, may have been overestimated (see below). 

Fig.~\ref{fig:cosmofit-euclidELG} is organized similarly and shows results for the \euclid\ sample. We see qualitatively similar results as for the \desi\ sample, with an excellent goodness-of-fit. The constraints on $f$ and \DaAP\ are similar to those for \desi, while the constraints on the length scales \sigv, $r_{\rm LP}$ and $r_{\rm ZC}$ are substantially broader. This is likely related to the lower value of the linear bias $b$ which decreases the significance of the BAO feature. We discuss this further below.

The possible overestimate of error (covariance) for the \desi\ sample mentioned above indicates that our analysis might benefit from the use of more accurate estimates of the covariance matrices than our rescaled Gauss-Poisson approximation described in Appendix~\ref{app:covariance}. Nevertheless, the results above form an important validation of the Zel'dovich smearing approximation, which is clearly capable of delivering unbiased cosmological constraints from samples with realistic nonlinearity. We discuss the constraining ability of the data in this model-agnostic framework in more detail in section~\ref{subsec:constraining-power} below.

\subsection{\emph{sdbmc} constraints}
\label{subsec:sdbmc-constraints}
\begin{figure}
\centering
\includegraphics[width=0.49\textwidth]{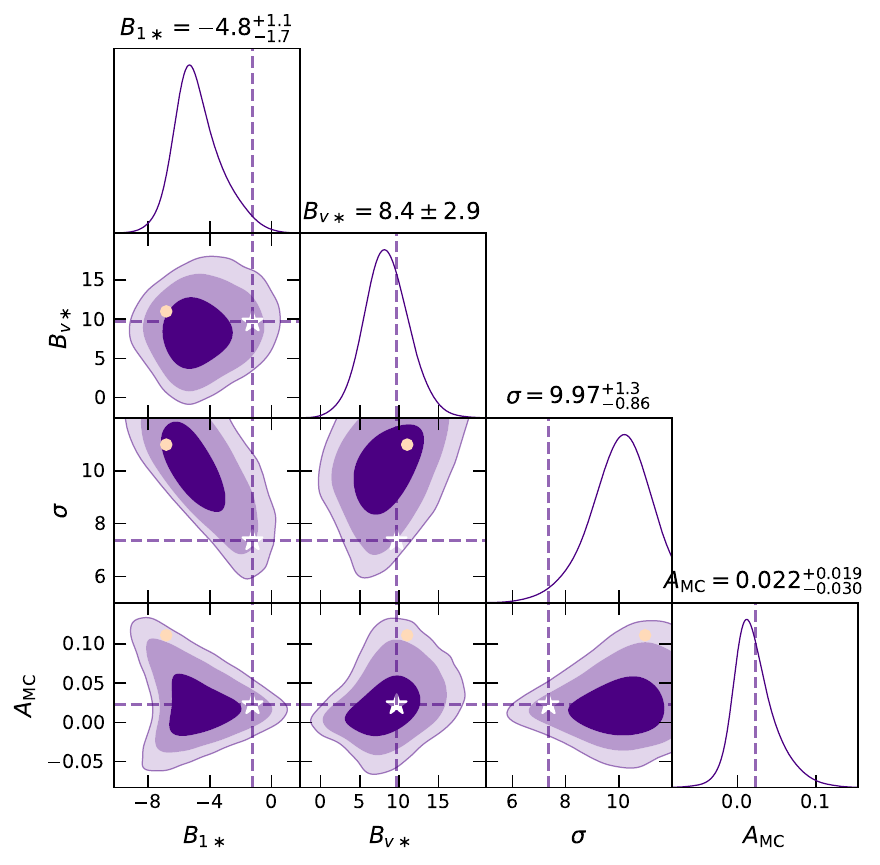}
\includegraphics[width=0.49\textwidth]{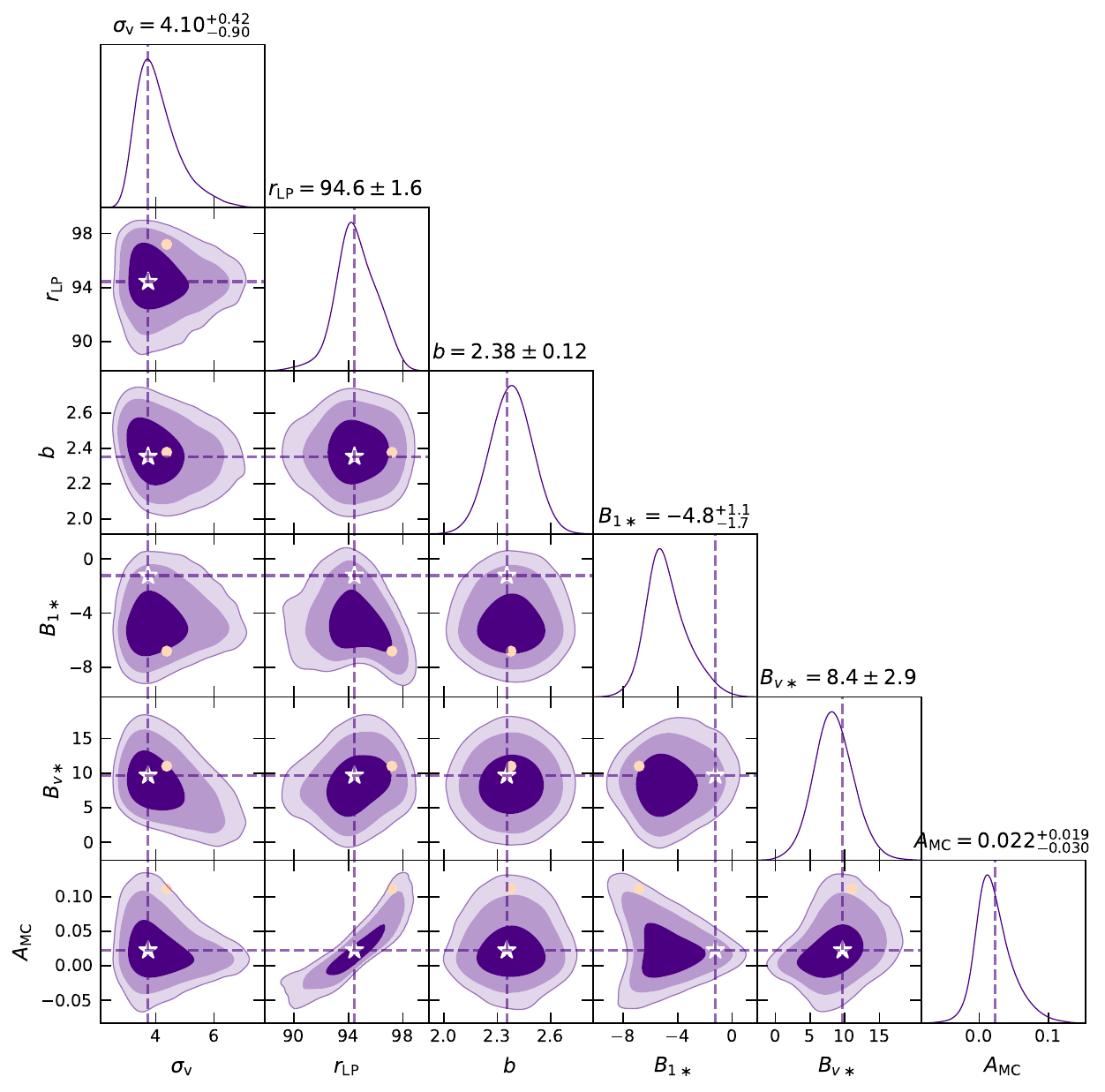}
\caption{Same as \emph{left panel} of Fig.~\ref{fig:cosmofit-desiLRG2} for the \desi\ sample, showing constraints on the \emph{sdbmc} parameters \emph{(left panel)} and a subset of cosmological and \emph{sdbmc} parameters \emph{(right panel)} to highlight some important degeneracies, notably $(A_{\rm MC},r_{\rm LP})$, $(B_{v\ast},\sigv)$. See text for a discussion and Fig.~\ref{fig:contours-desiLRG2} for joint constraints on all varied parameters.}
\label{fig:cosmo-sdbmc-desiLRG2}
\end{figure}

\begin{figure}
\centering
\includegraphics[width=0.49\textwidth]{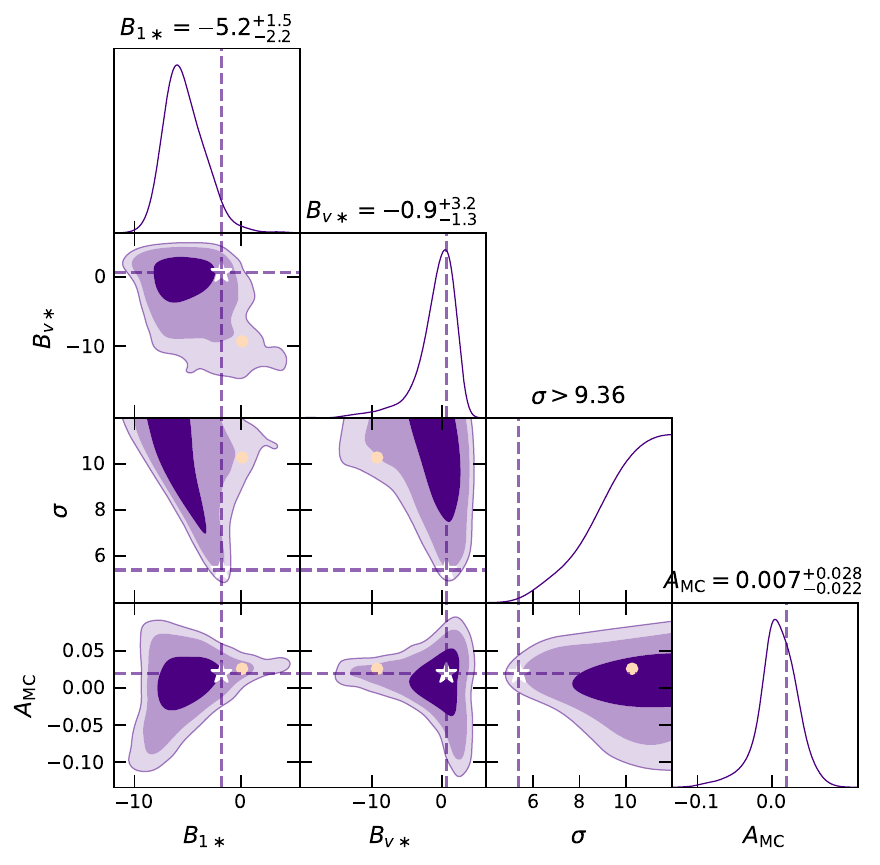}
\includegraphics[width=0.49\textwidth]{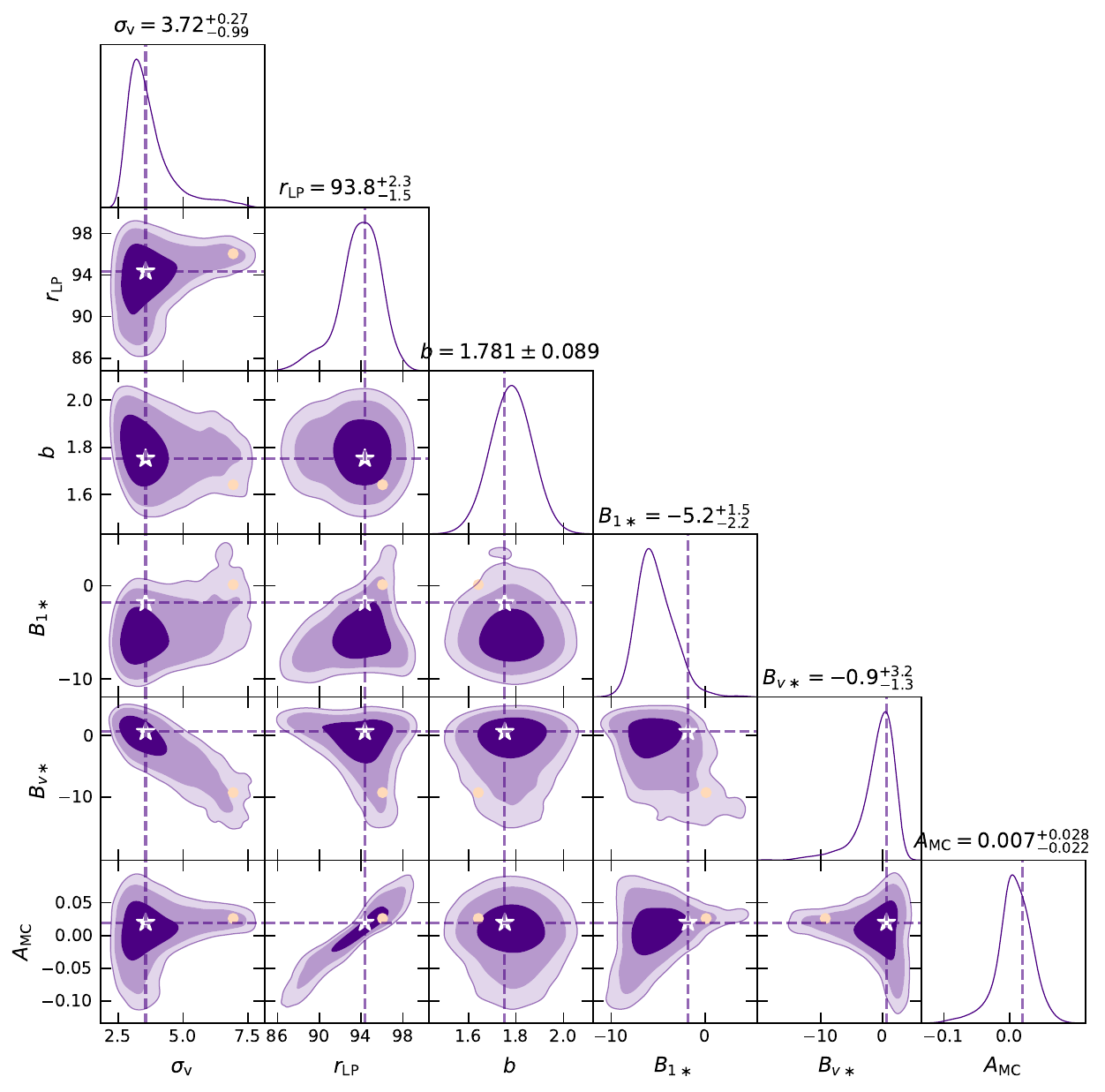}
\caption{Same as Fig.~\ref{fig:cosmo-sdbmc-desiLRG2}, for the \euclid\ sample.}
\label{fig:cosmo-sdbmc-euclidELG}
\end{figure}

Fig.~\ref{fig:cosmo-sdbmc-desiLRG2} shows the pairwise posteriors for the \emph{sdbmc} parameters and a combination of a subset of cosmological and \emph{sdbmc} parameters for the \desi\ sample. Fig.~\ref{fig:cosmo-sdbmc-euclidELG} is organized similarly and shows results for the \euclid\ sample. 
Unlike \citetalias{ps26a} who used a toy \emph{sdbmc} model, but similarly to \citetalias{ps25b} who analysed simulation data, for our \abacussummit\ samples we do not have ground truth values for the \emph{sdbmc} parameters, so we focus on the posterior distribution alone.

These constraints are largely similar to the toy model constraints of \citetalias{ps26a}. In particular, we see a pronounced degeneracy between $r_{\rm LP}$ and $A_{\rm MC}$ for both of our samples, which was also discussed by those authors. The degeneracy is especially strong in the \euclid\ sample, leading to a broad marginal distribution for $r_{\rm LP}$. As mentioned above, this is likely due to the lower value of the linear Eulerian bias $b$ for this sample as compared to the \desi\ sample. This leads to a lower signal-to-noise of the BAO feature, which allows for a larger horizontal leeway in predicting the location of the feature. Another interesting degeneracy is the one between $B_{v\ast}$ and \sigv, which leads to a broad tail in the marginal \sigv\ distribution, especially for the \euclid\ sample. This can again be traced back, at least partially, to the significance of the BAO feature which affects the recovery of $B_{v\ast}$ and \sigv\ through their common impact on the \emph{sdbmc} model (see equations~A.18-A.24 of \citetalias{ps26a}).

An interesting result in the \desi\ analysis is that the best fit value of the parameter $B_{v\ast}$ is \emph{positive}, being $\simeq+9.7$, with negative values being excluded with high significance. This was a concern in the analysis of \citetalias{ps25b} which indicated negative values based on halo and mock galaxy samples from the \texttt{HADES} \cite{hades} and \texttt{MINERVA} \cite{grieb2016} simulations, while a naive application of peaks theory would indicate that $B_{v\ast}$ should always be positive. In the present case, while the \desi\ analysis indicates a positive value, the \euclid\ sample demands a best fit close to zero, with a wide tail towards negative values. One hypothesis towards explaining negative values of $B_{v\ast}$, which was suggested by \citetalias{ps25b}, is that mass averaging over the scale-dependent effects from multiple narrow mass bins could mimic a negative $B_{v\ast}$. Another possibility is that \emph{second order} density bias effects might be as relevant as velocity bias effects at these scales, so that what we call $B_{v\ast}$ might encompass the sum total of these effects, which could then have an \emph{a priori} undetermined sign. It will be an interesting future exercise to theoretically segregate such effects in a model-agnostic framework; we choose not to do so here.

Finally, we also see that the constraint on the effective smearing scale $\sigma$ is affected by the upper end of its prior, mildly so for the \desi\ sample and much more strongly for the \euclid\ sample. As discussed by \citetalias{ps26a}, the upper limit of the $\sigma$ prior is decided by the calibration exercise done by \citetalias{ps25a} (see also Appendix~\ref{app:covariance}). As such, the fact that $\sigma$ is not well constrained (at least for \euclid) is a technical limitation of our analysis. It is reassuring, however, that the cosmological constraints (Figs.~\ref{fig:cosmofit-desiLRG2} and~\ref{fig:cosmofit-euclidELG}) are not biased as a result.

\subsection{Constraining ability of the model-agnostic framework}
\label{subsec:constraining-power}
In Appendix~\ref{app:allparams}, we show the pairwise posterior constraints on all sampled parameters in our model, as well as the derived parameters $\{f,r_{\rm peak},r_{\rm LP},r_{\rm ZC}\}$ for the \desi\ (Fig.~\ref{fig:contours-desiLRG2}) and \euclid\ sample (Fig.~\ref{fig:contours-euclidELG}). 

Correspondingly, we show a comparison between the posterior and prior for the cosmological parameters $\left\{\{w_m\},\DaAP\right\}$ for the \desi\ (Fig.~\ref{fig:contours-cosmoprior-desiLRG2}) and \euclid\ sample (Fig.~\ref{fig:contours-cosmoprior-euclidELG}). The comparisons show that not all of the cosmological parameters are well constrained by the data. E.g., for both samples, we see that although the coefficients $w_1,w_3,w_5,w_7,w_8$ are constrained tightly as compared to their priors, the remaining 4 coefficients are not, especially $w_6$ which is completely dominated by its prior in each case. As also discussed by \citetalias{ps26a}, while this suggests that our analysis might benefit from a further PCA-style reduction of the basis set, it remains desirable to retain the entire set and take advantage of its completeness and generalizability as demonstrated by \citetalias{ps25a}. We therefore choose to retain all the basis coefficients in our analysis, even though several of them are prior-dominated.

The behaviour of the constraints makes the comparison of our model-agnostic approach with traditional template-based approaches very interesting. For each analyzed sample, traditional BAO feature analyses typically rely on constraining the 2 parameters \DaAP\ and especially \Daiso\ (which is not included in our analysis) in order to recover cosmological information, while discarding $\gtrsim15$ parameters as being either cosmologically un-interesting or not robust enough.  Our analysis, on the other hand,  potentially extracts meaningful cosmological information from $\sim8$ parameters ($\{f,\sigv,w_1,w_3,w_5,w_7,w_8,\DaAP\}$ for, both, the \desi\ and \euclid\ samples) while entirely excluding the traditionally important \Daiso\ (see section~\ref{subsec:compare-traditional}), and additionally constrains the physically interpretable \emph{sdbmc} parameters. We emphasize that the exclusion of \Daiso\ in our analysis does not inherently degrade the quality of the outcome. This is because, unlike the traditional definition, \Daiso\ in our analysis does not include the sound horizon $r_{\rm s}$. Instead, the burden of information gleaned from the sound horizon is now carried by the basis coefficients $\{w_m\}$. Of course, it is natural to expect that cosmological constraints when interpreted in a specific model such as $\Lambda$CDM would be weaker in our approach than in a traditional approach, since the latter effectively imposes a strong prior on the shape of the 2pcf. We discuss this further in section~\ref{sec:conclude}.

\begin{figure}
\centering
\includegraphics[width=0.49\textwidth]{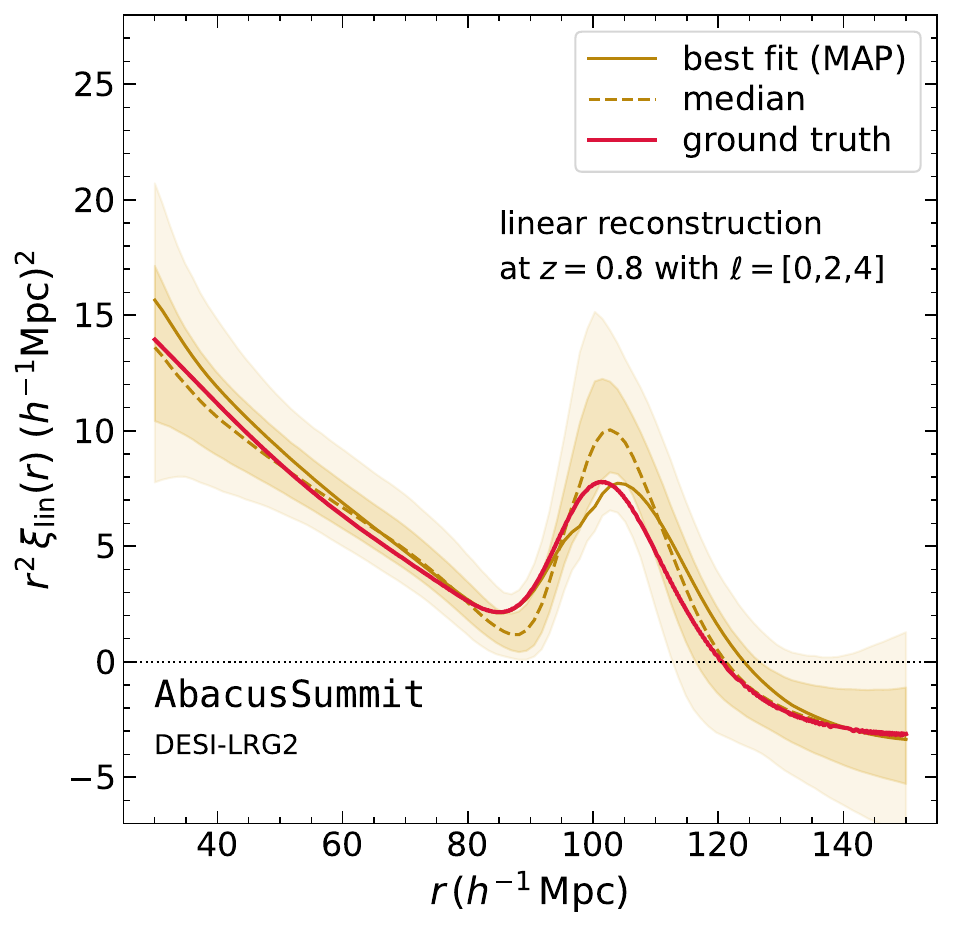}
\includegraphics[width=0.49\textwidth]{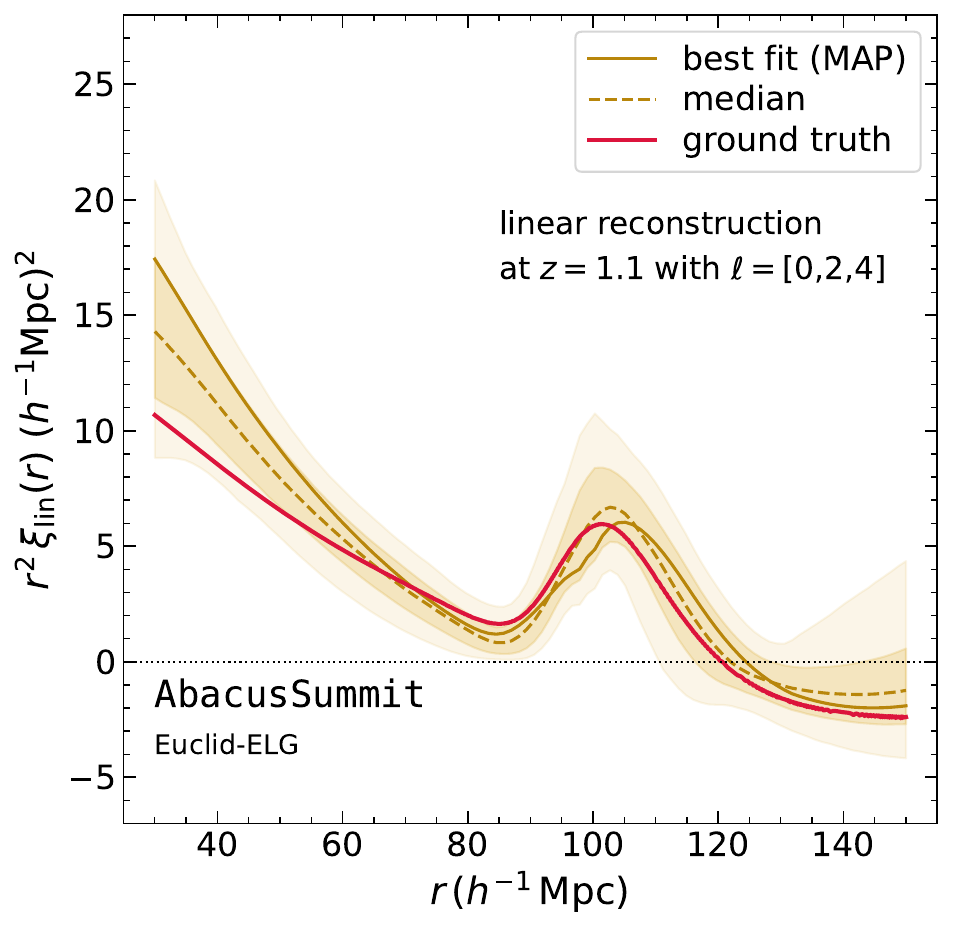}
\caption{
{\bf Reconstructed linear theory} for the \desi\ \emph{(left panel)} and \euclid\ sample \emph{(right panel)}. Red solid curve in each panel shows the ground truth for $\xi_{\rm lin}(r)$. Dashed curves with error bands show the respective median and central $68\%$ (dark bands) and $95\%$ confidence region (light bands) from the inference exercise, while the correspondingly coloured solid curve shows the best fit (MAP). We see good agreement between the best fit, median and ground truth, with the ground truth remaining inside the inferred $68\%$ interval over almost the entire range of scales for each sample. See text for a discussion.}
\label{fig:reconlin}
\end{figure}

\subsection{Reconstruction of linear 2pcf}
\label{subsec:reconlin}
Finally, as discussed by \citetalias{ps26a}, the posterior constraints on the basis coefficients $\{w_m\}$ and the linear Eulerian bias $b$ can be used to reconstruct the linear 2pcf $\xi_{\rm lin}(r)$ extrapolated to the epoch of the observed sample \cite{nsz21a,nsz21b,ps22}. Fig.~\ref{fig:reconlin} shows this reconstruction using our \desi\ and \euclid\ samples. 

We see good agreement between the distribution of the reconstructed $\xi_{\rm lin}(r)$ and the ground truth over the entire modelled range $30\leq r/(\Mpch) \leq 150$ in each case, thus extending the toy model results of \citetalias{ps26a} to semi-realistic tracer samples. As they noted, this means that the constraints on the $\{w_m\}$ could, in principle, contain substantially more cosmological information than is contained in the scales $r_{\rm LP}$ and $r_{\rm ZC}$. We note, though, that the constraints on $\xi_{\rm lin}(r)$ in each case are quite broad. This is a reflection of not only the degeneracies in the joint $\{w_m\}$ posterior but also the impact of opening up the \DaAP\ parameter space which is also mildly degenerate with the $\{w_m\}$ (see, e.g., the \emph{left panels} of Figs.~\ref{fig:cosmofit-desiLRG2}, \ref{fig:cosmofit-euclidELG}, and Figs.~\ref{fig:contours-cosmoprior-desiLRG2}, \ref{fig:contours-cosmoprior-euclidELG}). We return to this point in section~\ref{sec:conclude}.

\section{Discussion and Conclusion}
\label{sec:conclude}
In this work, we have completed the development of the Zel'dovich smearing approximation, which is a model-agnostic framework for cosmological inference using the BAO feature in redshift space. This framework was started in \citetalias{ps23} and subsequently augmented with two critical components: the \biseq\ basis for describing the linear 2pcf, developed in \citetalias{ps25a}, and the \emph{sdbmc} model to describe the impact of scale-dependent bias and mode coupling, developed in \citetalias{ps25b}. In \citetalias{ps26a}, these two ingredients were incorporated into the framework of \citetalias{ps23} and tested on toy data.  In the present work, we have (a) added a final ingredient that accounts for the effect of the inevitable fiducial cosmology that is used to convert observed angles and redshifts to comoving separations and (b) applied the resulting complete framework to perform inference on semi-realistic tracer samples constructed using halo catalogs from the high-resolution, large-volume \abacussummit\ simulation suite. The framework as a whole derives entirely from assuming the Zel'dovich approximation and basic ideas from peaks theory, without assuming any specific cosmological model such as $\Lambda$CDM. With the inclusion of the geometric ingredients described in this work, the model-agnostic Zel'dovich smearing framework is, in principle, ready for use on actual observational samples.
Although we have focused on its application to the large scale two-point function in galaxy surveys, currently acceptable models of two-point clustering in Ly-$\alpha$ datasets \cite{DESIDR2lya} are also amenable to our approach.  In addition, because the large scale three-point correlation function can be written as products of two point functions added to terms like those which we called mode-coupling \cite{PTreview}, our framework can be also be used to derive model-agnostic constraints from combinations of two- and three-point clustering.  We leave this for future work.

In the present context, the results (section~\ref{sec:results}) of the \abacussummit\ tracer samples are very encouraging. We have shown that the Zel'dovich smearing framework produces unbiased constraints on samples that mimic expected data from the DESI survey and \emph{Euclid} mission (the \desi\ and \euclid\ samples, respectively, described in section~\ref{subsec:catalogs}). It is worth noting, however, that the corresponding cosmological constraints on parameters such as the linear point $r_{\rm LP}$ represent larger errors ($\sim1.7\%$ for \desi\ and $\sim2.0\%$ for \euclid\ at $68\%$ confidence) than expected in previous model-agnostic work using the linear point (e.g., $\lesssim1.0\%$ as expected for comparable samples from the analysis in \cite{LPboss}). This is a consequence of two of the choices made in our setup: first, we explicitly model nonlinear effects such as smearing and \emph{sdbmc} which were not done in \cite{LPboss}\footnote{The argument in \cite{LPboss} and related work was that, the impact of these nonlinearities is a predictable shift of $0.5\%$ in $r_{\rm LP}$. However, this magnitude for the shift was derived from $\Lambda$CDM mocks and is therefore inherently model dependent. Our approach circumvents this by explicitly modelling the nonlinearity, albeit at the cost of an increased error in the $r_{\rm LP}$ estimate.} and, secondly, we have broken the connection between primordial physics (encoded by $\{w_m\}$) and late-time geometry (described by \DaAP; see also section~\ref{subsec:compare-traditional} for a discussion of why the related parameter \Daiso\ does not appear in our analysis). The resulting increase in error in the $r_{\rm LP}$ estimate is the price we pay for a fully model-agnostic setup. It will be very interesting to repeat such analyses using mock galaxy catalogs simulated in non-standard cosmologies, such as those presented by \cite{euclid26a}. 

It is also worth emphasizing that our analysis is completely up-front regarding the use of prior information (cf., the `weak $\Lambda$CDM prior' discussed in Appendix~\ref{app:priors}). The extent to which the large posterior widths discussed above are due to the width of the prior can therefore be easily assessed (e.g., by repeating the analysis with a tighter prior). In a traditional analysis, instead, the use of a $\Lambda$CDM template first `fixes' a sub-manifold of the agnostic parameter space, and a specification of a prior on the $\Lambda$CDM parameters then further constrains the volume accessible within this sub-manifold. Since the nature of this sub-manifold is not controllable,  adjusting the width of the $\Lambda$CDM prior itself does not fully capture the impact of the $\Lambda$CDM assumption (more on this below). 

The larger than expected error on $r_{\rm LP}$ naively suggests that our approach might be incapable of efficiently recovering cosmological information from a given tracer sample. One should keep in mind, however, that $r_{\rm LP}$ is not the only parameter that our approach constrains. As discussed in section~\ref{subsec:constraining-power}, we can expect the Zel'dovich smearing framework to extract useful cosmological information from $\sim8$ parameters, some of which, like the growth rate $f$ and smearing scale \sigv, are closely connected to traditional $\Lambda$CDM parameters like $\{\Omega_{\rm m},n_{\rm s},A_{\rm s}\}$. It is not clear \emph{a priori} how this information compares with, say, the recovery from a traditional template-based analysis. A proper comparison with a traditional analysis for $\Lambda$CDM would involve using comparable priors  and converting the model-agnostic posteriors presented in this work to $\Lambda$CDM posteriors on, say, $\{\Omega_{\rm m},n_{\rm s},A_{\rm s},h,\Omega_{\rm b},\Omega_{\rm k}\}$.

To construct a prior in the agnostic space that is comparable to one used for a traditional $\Lambda$CDM inference exercise, one can simply sample the $\Lambda$CDM prior and use it to generate a distribution in the agnostic space, treating the agnostic parameters as derived quantities in $\Lambda$CDM. The result would be a `strong $\Lambda$CDM' prior in the agnostic space, which essentially samples from the $\Lambda$CDM induced sub-manifold mentioned above. This is different than the `weak $\Lambda$CDM' prior described in Appendix~\ref{app:priors} and used in this work, where the correlations defining the sub-manifold were erased by only considering the marginal distributions one at a time. 
The conversion of posteriors from the agnostic space to $\Lambda$CDM would further need an `inverter' (implemented, say, as a neural network) that takes in our 13 model-agnostic cosmological parameters $\{f,\sigv,\{w_m\},\DaAP,f_{\rm v}\}$ and outputs the $\Lambda$CDM parameters. With an inverter of sufficiently high accuracy, the posterior chains for the model-agnostic parameters can be directly converted into a $\Lambda$CDM posterior. For consistency, the inverter must be trained only using the $\Lambda$CDM induced sub-manifold of the agnostic space; this can be easily achieved by sampling the inverter's training distribution using the same `strong $\Lambda$CDM' prior described above. Related ideas have been recently implemented by \cite{novell-masot+25-DESIPkBk,novell-masot+26} in DESI DR1 using the \emph{ShapeFit} \cite{shapefit2021} approach.

More generally, given its manner of construction, we might expect that the `weak $\Lambda$CDM' prior of this work is actually just a generic `weak' prior. This would be the case if, e.g., this prior also safely encloses any reasonable relations between the agnostic parameters induced by cosmologies different than $\Lambda$CDM. In that case, the analysis in this work would be generic, while the interpretation of the posteriors in the context of a specific model like $\Lambda$CDM would require convolving with the corresponding `strong' prior for that model. We are in the process of setting up such comparisons and will report the results in a future publication.

Finally, our model-agnostic approach has some interesting implications for the `Hubble tension' \cite{fm23}, which is the discrepancy between the value of the Hubble constant $H_0$ inferred from a low-redshift distance ladder involving Milky Way parallaxes, Cepheids and Type Ia supernovae \cite{rb24} and that inferred from high-redshift probes such as the cosmic microwave background \cite{Planck18-VI-cosmoparam} or an inverse distance ladder using the BAO feature \cite{camilleri+25}. Possible solutions to the tension (discounting observational systematics) have invoked a variety of modifications to standard cosmology at early as well as late epochs (see, e.g., \cite{khalife+24} and references therein). Our Zel'dovich smearing approximation for describing the BAO feature explicitly separates the impact of primordial physics (captured by the basis coefficients $\{w_m\}$, the linear growth rate $f$ and the smearing scale \sigv) from the effects of late-time geometry (captured by \DaAP). For comparison, in the \emph{ShapeFit} approach, this separation can be achieved by marginalising over the sound horizon; see the `sound horizon-free' $H_0$ estimates of \cite{novell-masot+26}. In our case, by comparing the posterior constraints on the agnostic parameters under the `weak' and `strong $\Lambda$CDM' priors discussed above, it might be possible to isolate the $H_0$ discrepancy as arising from either the primordial or late-time sector. E.g., an analysis using the `strong $\Lambda$CDM' prior would be expected to be consistent with existing BAO inverse distance ladder estimates of $H_0$, while an analysis using the `weak' prior might yield valuable insights by identifying which set of agnostic parameters shows the largest discrepancy, if any.  We will pursue this idea in future work.

\section*{Data availability}
All the code used for producing the measurements, MCMC analysis and plots in this work is publicly available in the repository \url{https://github.com/a-paranjape/zeldovich-smearing}. The same code can also be used to reproduce the results of \citetalias{ps26a}, and is expected to be useful for future analyses of observational samples. We provide convenient Jupyter notebooks to explore the theoretical model and perform MCMC analyses. The repository also contains the 2pcf and power spectrum multipole measurements for the \desi\ and \euclid\ samples defined using the \abacussummit\ halo catalogs as described in the text (see section~\ref{subsec:catalogs} and Figs.~\ref{fig:pwise-desiLRG2} and~\ref{fig:pwise-euclidELG}), along with the scaled, smoothed Gauss-Poisson covariance matrices used for each iteration of each sample (see Appendix~\ref{app:covariance} and Fig.~\ref{fig:GPcov}). The MCMC chains generated in this work, being heavier than the size limitations of the repository, are available upon reasonable request to AP. The use of any of these data products or code may please cite \citetalias{ps26a}, the present paper and the above repository.

\section*{Acknowledgments}
We are grateful to the \abacussummit\ team for making their data products publicly available.
AP thanks Shadab Alam for help with accessing the \abacussummit\ halo catalogs. 
We gratefully acknowledge the use of high performance computing facilities at TIFR, Mumbai and IUCAA, Pune.
The research of AP is supported by the Associates Scheme of ICTP, Trieste.  
RKS is grateful to the IUCAA community for their hospitality in Spring 2026.
This work made extensive use of the open source computing packages NumPy \citep{vanderwalt-numpy},\footnote{\url{http://www.numpy.org}} SciPy \citep{scipy},\footnote{\url{http://www.scipy.org}} Matplotlib \citep{hunter07_matplotlib},\footnote{\url{https://matplotlib.org/}} AstroPy \cite{astropy:2013,astropy:2018,astropy:2022}\footnote{\url{https://docs.astropy.org/en/stable/index.html}} and Jupyter Notebook.\footnote{\url{https://jupyter.org}} 

\bibliography{references}
\appendix

\section{Details of model construction}
\label{app:model}

\subsection{Useful relations}
We will use the following integral relations obeyed by the Legendre polynomials,
\begin{align}
\int_{-1}^1\frac{\der\mu}{2}\,\Pell{\ell}(\mu)\,\Pell{\ell^\prime}(\mu) &= \frac{1}{(2\ell+1)}\,\delta_{\ell\ell^\prime}\,,
\label{eq:Legendre-orthog}\\
\int_{-1}^1\frac{\der\mu}{2}\,\Pell{\ell}(\mu)\,\Pell{\ell^\prime}(\mu)\,\Pell{\ell^{\prime\prime}}(\mu) &= 
\tjred{\ell}{\ell^\prime}{\ell^{\prime\prime}}^2
\,,
\label{eq:Legendre-3j}
\end{align}
where the first relation reflects orthogonality and $\tjs{j_1}{j_2}{j_3}{m_1}{m_2}{m_3}$ is a Wigner $3j$ symbol \citep{Wigner1993}. We will also  use the identity\footnote{This can be derived using the generating function of the Legendre polynomials, $\left(1-2t\mu+t^2\right)^{-1/2}=\sum_{\ell=0}^\infty\,t^\ell\,\Pell{\ell}(\mu)$, along with the orthogonality relation of the $\Pell{\ell}$. For a proof, see \url{https://math.stackexchange.com/questions/1586202/monomials-in-terms-of-legendre-polynomials}.}
\beq
\mu^{2n} = \sum_{k=0}^n\frac{(2n)!}{2^kk!}\,\frac{\left(4(n-k)+1\right)}{\left(4n-2k+1\right)!!}\,\Cal{P}_{2(n-k)}(\mu)\,,
\label{eq:mu^2n-Pell-identity}
\eeq

\subsection{Configuration space 2pcf multipoles}
\label{subapp:2pcf}
The theory predicts (or the Universe produces) $\xi(\ss_{\rm t})$, which we then  convert to $\xi_{\rm obs}(\ss_{\rm f})$. When the $\alpha$'s are all close to unity, these can be related using (e.g., \cite{xu+13})
\begin{align}
\xi_{\rm obs}(\ss_{\rm f}) &= \xi(\ss_{\rm t}(\ss_{\rm f};\boldsymbol{\alpha})) \notag\\
&= \xi(\ss_{\rm f})
+ (\boldsymbol{\alpha}-\boldsymbol{\alpha}_0) \cdot \left.\nabla_{\boldsymbol{\alpha}}\xi(\ss_{\rm t}(\ss_{\rm f};\boldsymbol{\alpha})) \right|_{\boldsymbol{\alpha}=\boldsymbol{\alpha}_0} + \ldots \notag\\
&= \xi(\ss_{\rm f}) + \Delta\boldsymbol{\alpha} \cdot \left. \left(\nabla_{\boldsymbol{\alpha}}\,\ss_{\rm t} \cdot \nabla_{\ss_{\rm t}}\xi(\ss_{\rm t}) \right)\right|_{\boldsymbol{\alpha}=\boldsymbol{\alpha}_0} + \ldots \,,
\label{eq:xiobs<->xi}
\end{align}
where we denoted 
\beq
\boldsymbol{\alpha}=(\aAP,\aperp)\,; \quad \boldsymbol{\alpha}_0=(1,1)\,; \quad \Delta\boldsymbol{\alpha}\equiv\boldsymbol{\alpha}-\boldsymbol{\alpha}_0\,,
\label{eq:Delta-alpha-def}
\eeq
Taylor expanded to leading order in $\Delta\boldsymbol{\alpha}$ and applied the chain rule. In this notation, setting $\Delta\boldsymbol{\alpha}=0$ or $\boldsymbol{\alpha}=\boldsymbol{\alpha}_0$ is the same as setting $\ss_{\rm t}=\ss_{\rm f}$.

We interpret the vector $\ss_{\rm t}$ as $(s_{\rm t},\mu_{s{\rm t}})$, using which the second term in the last equality of \eqn{eq:xiobs<->xi} can be manipulated as
\begin{align}
\Delta\boldsymbol{\alpha} \cdot \left. \left(\nabla_{\boldsymbol{\alpha}}\,\ss_{\rm t} \cdot \nabla_{\ss_{\rm t}}\xi(\ss_{\rm t}) \right)\right|_{\boldsymbol{\alpha}=\boldsymbol{\alpha}_0}
&= \p_{s_{\rm t}}\xi\left[\DaAP\p_{\aAP}s_{\rm t} + \Delta\aperp\p_{\aperp}s_{\rm t} \right] \notag\\
&\ph{\p_s\xi}
+ \p_{\mu_{s{\rm t}}}\xi\left[\DaAP\p_{\aAP}\mu_{s{\rm t}} + \Delta\aperp \p_{\aperp}\mu_{s{\rm t}}\right] \,,
\end{align}
with the understanding that all derivatives are to be evaluated at $\Delta\boldsymbol{\alpha}=0$. The matrix $\nabla_{\boldsymbol{\alpha}}\,\ss_{\rm t}$ in this case becomes
\beq
\nabla_{\boldsymbol{\alpha}}\,\ss_{\rm t} = 
\begin{pmatrix}
\p_{\aAP}s_{\rm t} & \p_{\aAP}\mu_{s{\rm t}} \\
\p_{\aperp}s_{\rm t} & \p_{\aperp}\mu_{s{\rm t}}
\end{pmatrix}
= 
\begin{pmatrix}
\mu_{s{\rm f}}^2\,s_{\rm f} & \mu_{s{\rm f}}\left(1-\mu_{s{\rm f}}^2\right) \\
-s_{\rm f} & 0
\end{pmatrix} \,,
\eeq
and we can write
\begin{align}
\Delta\boldsymbol{\alpha} \cdot \left. \left(\nabla_{\boldsymbol{\alpha}}\,\ss_{\rm t} \cdot \nabla_{\ss_{\rm t}}\xi(\ss_{\rm t}) \right)\right|_{\boldsymbol{\alpha}=\boldsymbol{\alpha}_0}
&= s_{\rm f}\p_{s_{\rm f}}\xi(\ss_{\rm f})\left[\left(\frac{\DaAP}{3} - \Delta\aperp \right) + \frac23\DaAP\,\Pell{2}(\mu_{s{\rm f}}) \right] \notag\\
&\ph{s_{\rm f}\p_{s_{\rm f}}\xi} 
+ \DaAP(1-\mu_{s{\rm f}}^2) \mu_{s{\rm f}}\,\p_{\mu_{s{\rm f}}}\xi(\ss_{\rm f})\,,
\end{align}
where we used \eqn{eq:mu^2n-Pell-identity} to write $\mu^2=(2\Pell{2}(\mu)+1)/3$.
A harmonic transform then gives
\begin{align}
\xiellobs{\ell}(s_{\rm f}) &= \left(2\ell+1\right)\int_{-1}^1\frac{\der\mu_{s{\rm f}}}{2}\,\Pell{\ell}(\mu_{s{\rm f}})\,\xi_{\rm NL,obs}(\ss_{\rm f}) \notag\\
&= \xiell{\ell}(s_{\rm f}) - \left(\Delta\aperp-\frac{\DaAP}{3}\right) s_{\rm f}\p_{s_{\rm f}}\,\xiell{\ell}(s_{\rm f}) \notag\\
&\ph{\xiell{}}
+ \DaAP\left(2\ell+1\right) \int_{-1}^1\frac{\der\mu_{s{\rm f}}}{2}\,\Pell{\ell}(\mu_{s{\rm f}})\bigg[ \frac23\Pell{2}(\mu_{s{\rm f}}) s_{\rm f}\p_{s_{\rm f}}\,\xi(\ss_{\rm f}) \notag\\ 
&\ph{\xiell{+ \DaAP\left(2\ell+1\right) \int_{-1}^1\frac{\der\mu_{s{\rm f}}}{2}}}
+ \mu_{s{\rm f}}(1-\mu_{s{\rm f}}^2)\,\p_{\mu_{s{\rm f}}}\,\xi(\ss_{\rm f}) \bigg] + \ldots
\label{eq:xiobs(ell)-temp}
\end{align}
It is useful to note that, if we work with $(\aAP,\aiso)$ instead of the scale variables $(\aAP,\aperp)$, where $\aiso$ was defined in \eqn{eq:alpha_iso-def}, then we can write 
\beq
\Delta\aperp-\frac{\DaAP}{3} = \Daiso\,.
\eeq
To proceed further, we harmonic expand $\xi(\ss_{\rm f})$ under the integral in \eqn{eq:xiobs(ell)-temp} as
\beq
\xi(\ss_{\rm f}) = \sum_{\ell^\prime\geq0} \xiell{\ell^\prime}(s_{\rm f})\,\Cal{P}_{\ell^\prime}(\mu_{s{\rm f}}) \simeq
\sum_{\ell^\prime=0,2,4} \xiell{\ell^\prime}(s_{\rm f})\,\Cal{P}_{\ell^\prime}(\mu_{s{\rm f}}) \,,
\eeq
where we recognize that the nonlinearly evolved 2pcf only depends on even powers of the cosine angle, so that only even multipoles contribute, and \emph{we ignore multipoles higher than $\ell^\prime=4$ since they are expected to contribute negligible signal-to-noise.} The first term in square brackets in \eqn{eq:xiobs(ell)-temp} can then be written in terms of a symmetric matrix $\Cal{C}_{\ell\ell^\prime}$ derived from the Wigner $3j$ symbols (equations 23, 24 of \citetalias{ps23}; see equation~\ref{eq:Legendre-3j}): 
\beq
\Cal{C}_{\ell\ell^\prime} \equiv \tjred{\ell}{\ell^\prime}{2}^2 = 
\begin{pmatrix}
\Cal{C}_{00} & \Cal{C}_{02} & \Cal{C}_{04} \\
\Cal{C}_{20} & \Cal{C}_{22} & \Cal{C}_{24} \\
\Cal{C}_{40} & \Cal{C}_{42} & \Cal{C}_{44} \end{pmatrix}
=
\begin{pmatrix}
0 & 1/5 & 0 \\
1/5 & 2/35 & 2/35 \\
0 & 2/35 & 20/693 
\end{pmatrix}\,.
\label{eq:Cmatrix}
\eeq
The second term can be simplified after writing the powers of $\mu_{s{\rm f}}$ in terms of Legendre polynomials using the identity \eqref{eq:mu^2n-Pell-identity}. The result can be organized in terms of a non-symmetric matrix $\Cal{A}_{\ell\ell^\prime}$ defined as
\beq
\Cal{A}_{\ell\ell^\prime} = 
\begin{pmatrix}
\Cal{A}_{00} & \Cal{A}_{02} & \Cal{A}_{04} \\
\Cal{A}_{20} & \Cal{A}_{22} & \Cal{A}_{24} \\
\Cal{A}_{40} & \Cal{A}_{42} & \Cal{A}_{44} \end{pmatrix}
=
\begin{pmatrix}
0 & 2/5 & 0 \\
0 & 2/7 & 20/21 \\
0 & -24/35 & 20/77 
\end{pmatrix}\,.
\label{eq:Amatrix}
\eeq
This leads to (cf., equation~20 of  \cite{xu+13})\footnote{We excluded writing a term $-\frac{40}{33}\xiell{4}(s_{\rm f})\,\delta_{\ell,6}$ inside the square brackets in \eqn{eq:xiobs(ell)-sbins} for consistency with our assumption that we only consider the multipoles $\ell=0,2,4$.}
\begin{align}
\xiellobs{\ell}(s_{\rm f}) &= \xiell{\ell}(s_{\rm f}) - \Daiso\,s_{\rm f}\p_{s_{\rm f}}\xiell{\ell}(s_{\rm f}) \notag\\
&\ph{\xiell{}}
+ \DaAP \bigg[\frac23\left(2\ell+1\right) \sum_{\ell^\prime}\Cal{C}_{\ell\ell^\prime}\,s_{\rm f}\p_{s_{\rm f}}\xiell{\ell^\prime}(s_{\rm f}) + \sum_{\ell^\prime}\Cal{A}_{\ell\ell^\prime}\, \xiell{\ell^\prime}(s_{\rm f}) \bigg] \,.
\label{eq:xiobs(ell)-sbins}
\end{align}
To mitigate the degeneracies discussed in section~\ref{subsec:compare-traditional} that arise due to the term proportional to \Daiso, we absorb the effect of this term into the Zel'dovich smearing model itself, specifically into the basis coefficients $\{w_m\}$. 

This can be achieved by working with the scaled variables $\{y_{\rm t},y_{\rm f}\}$ instead of $\{s_{\rm t},s_{\rm f}\}$, where
\beq
y_{\rm f} \equiv s_{\rm f}/D_{\rm Vf} \,;\quad y_{\rm t} \equiv s_{\rm t}/D_{\rm V} = s_{\rm t}\,\aiso/D_{\rm Vf}\,,
\label{eq:y,yf-def}
\eeq
with $D_{\rm V}$ being the isotropized comoving distance measure defined in \eqn{eq:DV-def}. 
To see why this choice is convenient, consider the analog of \eqref{eq:s(sf,musf)} for $y_{\rm f},y_{\rm t}$:
\beq
\frac{y_{\rm f}}{y_{\rm t}} = \frac{s_{\rm f}/s_{\rm t}}{\aiso} 
= \frac{\alpha_{\rm AP}^{1/3}}{\sqrt{1 + \mu_{\rm sf}^2(\alpha_{\rm AP}^2-1)}}\,,
\label{eq:y(yf,muf)}
\eeq
which does not depend on \aiso.
The relation between $\mu_{s{\rm t}}$ and $\mu_{s{\rm f}}$ is still given by \eqn{eq:mus(sf,musf)}, which is also independent of \aiso.
This means that, when transforming $\yy$ (rather than \ss) from the true to the fiducial cosmology, the dependence on \aiso\ has been removed and only \aAP\ enters. 

To leverage this behaviour, we \emph{assume} the relation 
\beq
\xi_{\rm obs}(\yy_{\rm f}) = \xi(\yy_{\rm t}(\yy_{\rm f};\boldsymbol{\alpha})) \,,
\eeq
instead of the first line of \eqref{eq:xiobs<->xi}. This is always possible in any model, with the understanding that all physical length scales $r$ in the model should also be scaled to $r/D_{\rm V}$. It is then straightforward to repeat the previous calculation,
approximating 
\begin{align}
\yy_{\rm t} &= \frac{1}{D_{\rm V}}\,\ss_{\rm t}(\ss_{\rm f};\boldsymbol{\alpha}) \notag\\
&= \frac{\aiso}{D_{\rm Vf}}\left(\ss_{\rm f} + \left(\Delta\boldsymbol{\alpha}\cdot\nabla_{\boldsymbol{\alpha}}\ss_{\rm t}\right)|_{\Delta\boldsymbol{\alpha}=0} + \ldots\right)\, \notag\\
&= \alpha_{\rm iso}\left(\yy_{\rm f}+ D_{\rm Vf}^{-1}\left(\Delta\boldsymbol{\alpha}\cdot\nabla_{\boldsymbol{\alpha}}\ss_{\rm t}\right)|_{\Delta\boldsymbol{\alpha}=0} + \ldots\right) \,, 
\end{align}
which leads to
\begin{align}
\xi_{\rm obs}(\yy_{\rm f}) &=  \xi\left(\aiso\left[\yy_{\rm f}+ D_{\rm Vf}^{-1}\left(\Delta\boldsymbol{\alpha}\cdot\nabla_{\boldsymbol{\alpha}}\ss_{\rm t}\right)|_{\Delta\boldsymbol{\alpha}=0}\right]\right) \notag\\
&= \xi\left(\aiso\yy_{\rm f}\right) + \frac{\Delta\boldsymbol{\alpha}\cdot\nabla_{\boldsymbol{\alpha}}\ss_{\rm t}}{D_{\rm Vf}}\cdot\nabla_{\yy}\xi\,\aiso + \Cal{O}\left(\Delta\alpha^2\right) \notag\\
&=  \xi\left(\yy_{\rm f} + \Daiso\yy_{\rm f}\right) + \Delta\boldsymbol{\alpha} \cdot \left. \left(\nabla_{\boldsymbol{\alpha}}\,\ss_{\rm t} \cdot \nabla_{\ss_{\rm t}}\xi(\ss_{\rm t}) \right)\right|_{\boldsymbol{\alpha}=\boldsymbol{\alpha}_0} + \ldots\notag\\
&=  \xi(\yy_{\rm f}) + \Daiso\,s_{\rm t}\,\p_{s{\rm t}}\xi|_{\boldsymbol{\alpha}=\boldsymbol{\alpha}_0} + \Delta\boldsymbol{\alpha} \cdot \left. \left(\nabla_{\boldsymbol{\alpha}}\,\ss_{\rm t} \cdot \nabla_{\ss_{\rm t}}\xi(\ss_{\rm t}) \right)\right|_{\boldsymbol{\alpha}=\boldsymbol{\alpha}_0} + \ldots \,,
\label{eq:xiobs-ybins}
\end{align}
where we ignored contributions of $\Cal{O}(\Delta\alpha^2)$ in all terms. The last term in \eqn{eq:xiobs-ybins} was already calculated in the derivation leading to \eqn{eq:xiobs(ell)-sbins}. Importantly, as expected, the term involving \Daiso\ in \eqn{eq:xiobs-ybins} \emph{precisely cancels} the corresponding contribution from the last term (cf., equation~\ref{eq:xiobs(ell)-sbins}), leaving
\begin{align}
\xiellobs{\ell}(y_{\rm f}) &= \xiell{\ell}(y_{\rm f}) + \DaAP \bigg[\frac23\left(2\ell+1\right) \sum_{\ell^\prime}\Cal{C}_{\ell\ell^\prime}\,y_{\rm f}\p_{y_{\rm f}}\xiell{\ell^\prime}(y_{\rm f}) + \sum_{\ell^\prime}\Cal{A}_{\ell\ell^\prime}\, \xiell{\ell^\prime}(y_{\rm f}) \bigg] \,.
\label{eq:xiobs(ell)-ybins}
\end{align}
In fact, as \eqn{eq:y(yf,muf)} shows, this independence on \Daiso\ is exact, and not restricted to the linear order terms.

For convenience, when presenting results, we simply multiply $y_{\rm f}$ with the known value of $D_{\rm Vf}$, so that the final model prediction can be written as in \eqn{eq:xiobs(ell)}. Correspondingly, we then compare all primordial length scales $r$ predicted by the model with their scaled ground truth values $r_{\rm true}\times\alpha_{\rm iso,true}$.

\subsection{Power spectrum multipole integrals}
\label{subapp:Pk}
To understand the expected effect of the fiducial cosmology on the power spectrum multipole integrals, it is useful to start with the continuum form of the estimator \eqref{eq:Sig2ell-estimate}. If we knew the true cosmology, we could write this as
\beq
\Sigell{\ell} = \frac{1}{6\pi^2}\int_{k_{\rm min}}^{k_{\rm max}}\der k\,P^{(\ell)}(k) = \frac13\int_{k_{\rm min}}^{k_{\rm max}}\der\ln k\,k^{-2}\,\Dellsq{\ell}(k)\,,
\label{eq:Sig2ell-truth-def}
\eeq
where $k_{\rm min}$ and $k_{\rm max}$ are comoving scales in the ground truth cosmology and $\Dellsq{\ell}(k)$ is the inverse Hankel transform of $\xiell{\ell}(s)$, 
\beq
\Dellsq{\ell}(k) = \frac{2k^3}{\pi}\,(-i)^\ell\int_0^\infty\der s\,s^2\,j_\ell(ks)\,\xiell{\ell}(s)\,,
\label{eq:Dell-def}
\eeq
so that
\beq
\xiell{\ell}(s) = i^\ell\int\der\ln k\,\Dellsq{\ell}(k)\,j_\ell(ks)\,.
\label{eq:Hankel}
\eeq
The more realistic observed quantities, in the case where the fiducial cosmology differs from the true one, are then
\beq
\Sigellobs{\ell} = \frac13\int_{k_{\rm min,f}}^{k_{\rm max,f}}\der\ln k_{\rm f}\,k_{\rm f}^{-2}\,\Dellsqobs{\ell}(k_{\rm f})\,,
\label{eq:Sig2ell-obs-def}
\eeq
where $k_{\rm min/max,f} \equiv k_{\rm min/max}\,\aiso^{-1}$ are the corresponding bounding scales in the fiducial cosmology. This can be manipulated as
\begin{align}
\Sigellobs{\ell} &= \frac{2(-i)^\ell}{3\pi}\int_{k_{\rm min,f}}^{k_{\rm max,f}}\der k_{\rm f} \int_0^\infty\der s_{\rm f}\,s_{\rm f}^2\,\xiellobs{\ell}(s_{\rm f})\,j_\ell(k_{\rm f}s_{\rm f})
\notag\\
&= D_{\rm Vf}^2 \frac{2(-i)^\ell}{3\pi}\int_0^\infty\der y_{\rm f}\,y_{\rm f}\,\xiellobs{\ell}(y_{\rm f}) \int_{\kappa_{\rm min}y_{\rm f}}^{\kappa_{\rm max}y_{\rm f}}\der x\,j_\ell(x) \notag\\
&= D_{\rm Vf}^2 \frac{2(-i)^\ell}{3\pi}\int_0^\infty\der y\,y\,\xiellobs{\ell}(y) \Cal{J}_\ell\left(\kappa_{\rm min}y,\kappa_{\rm max}y\right) \,,
\label{eq:Sig2ell-obs-temp1}
\end{align}
where we switched from $s_{\rm f}\to y_{\rm f}$, introduced 
$\kappa_{\rm min/max}\equiv k_{\rm min/max,f}D_{\rm Vf}=k_{\rm min/max}D_{\rm V}$ and, in the last line, wrote the dummy integration variable as $y$ instead of $y_{\rm f}$ and defined the function
\beq
\Cal{J}_\ell(x_{\rm min},x_{\rm max}) \equiv \int_{x_{\rm min}}^{x_{\rm max}}\der x\,j_\ell(x)\,.
\label{eq:Jell-def}
\eeq
Replacing $\xiellobs{\ell}(y)$ under the integral in \eqref{eq:Sig2ell-obs-temp1} with the expression in \eqref{eq:xiobs(ell)-ybins} leads to terms involving the derivative of $y^2\Cal{J}_\ell$, which can be simplified as
\begin{align}
\p_y\left(y^2\Cal{J_\ell}\right) &= 2y\Cal{J}_\ell\left(\kappa_{\rm min}y,\kappa_{\rm max}y\right) + y^2\p_y\int_{\kappa_{\rm min}y}^{\kappa_{\rm max}y}\der x\,j_\ell(x) \notag\\
&= 2y\Cal{J}_\ell\left(\kappa_{\rm min}y,\kappa_{\rm max}y\right) + y^2\left[\kappa_{\rm max}j_\ell(\kappa_{\rm max}y) - \kappa_{\rm min}j_\ell(\kappa_{\rm min}y)\right]\,.
\end{align}
Using \eqn{eq:xiobs(ell)-ybins} in \eqn{eq:Sig2ell-obs-temp1}, we have
\begin{align}
\frac{\Sigellobs{\ell}}{D_{\rm Vf}^2} &= \frac{2(-i)^\ell}{3\pi}\int_0^\infty \der y\,y\,\Cal{J}_\ell \left[\xiell{\ell}(y) + \DaAP \sum_{\ell^\prime}\Cal{A}_{\ell\ell^\prime}\xiell{\ell^\prime}(y) \right] \notag\\
&\ph{\int\der} 
+ \frac{2(-i)^\ell}{3\pi}\int_0^\infty \der y\,\p_y\left(y^2\Cal{J}_\ell\right) \left[
-\frac23(2\ell+1)\DaAP\sum_{\ell^\prime}\Cal{C}_{\ell\ell^\prime}\xiell{\ell^\prime}(y) \right] \notag\\
&= \frac{2(-i)^\ell}{3\pi}\int_0^\infty \der y\,\xiell{\ell}(y) \,y\,\Cal{J}_\ell\left(\kappa_{\rm min}y,\kappa_{\rm max}y\right) 
\notag\\
&\ph{\int\der}
+\DaAP\,\frac{2(-i)^\ell}{3\pi} \sum_{\ell^\prime}\int_0^\infty \der y\,\xiell{\ell^\prime}(y) \bigg[y\Cal{J}_\ell \left(A_{\ell\ell^\prime} - \frac43(2\ell+1)\Cal{C}_{\ell\ell^\prime}\right) \notag\\
&\ph{+\DaAP\,\frac{2(-i)^\ell}{3\pi} \sum_{\ell^\prime}}
- \frac23(2\ell+1)\Cal{C}_{\ell\ell^\prime} \,y^2\left(\kappa_{\rm max}j_\ell(\kappa_{\rm max}y) - \kappa_{\rm min}j_\ell(\kappa_{\rm min}y)\right)\bigg] \,.
\label{eq:Sig2ell-obs-temp2}
\end{align}
We can manipulate \eqn{eq:Sig2ell-obs-temp2} in three ways. First, we recognize 
\begin{align}
\Sigell{\ell} &= \frac{2(-i)^\ell}{3\pi}\int_0^\infty\der s\,s\,\xiell{\ell}(s)\Cal{J}_\ell\left(k_{\rm min}s,k_{\rm max}s\right)\notag\\
&= D_{\rm V}^2\, \frac{2(-i)^\ell}{3\pi}\int_0^\infty\der y\,y\,\xiell{\ell}(y)\Cal{J}_\ell\left(\kappa_{\rm min}y,\kappa_{\rm max}y\right)\,,
\end{align}
(see equations~\ref{eq:Sig2ell-truth-def}, \ref{eq:Sig2ell-obs-def} and~\ref{eq:Sig2ell-obs-temp1}).
Next, we use \eqn{eq:Dell-def} to write
\begin{align}
&D_{\rm V}^2\frac{2(-i)^\ell}{3\pi}\int_0^\infty \der y\,y^2\,\xiell{\ell}(y) \left(\kappa_{\rm max}j_\ell(\kappa_{\rm max}y) - \kappa_{\rm min}j_\ell(\kappa_{\rm min}y)\right) \notag\\
&=\frac{2(-i)^\ell}{3\pi}\int_0^\infty \der s\,s^2\,\xiell{\ell}(s) \left(k_{\rm max}j_\ell(k_{\rm max}s) - k_{\rm min}j_\ell(k_{\rm min}s)\right) \notag\\
&\ph{\int\der s}
= \frac13 \left[k_{\rm max}^{-2}\Dellsq{\ell}(k_{\rm max}) - k_{\rm min}^{-2}\Dellsq{\ell}(k_{\rm min}) \right] \notag\\
&\ph{\int\der s}
= \frac13 \int_{k_{\rm min}}^{k_{\rm max}} \frac{\der k}{k}\,\frac{\p}{\p\ln k} \left(k^{-2}\Dellsq{\ell}(k) \right) \notag\\
&\ph{\int\der s}
= \frac13 \int_{k_{\rm min}}^{k_{\rm max}} \der\ln k\,k^{-2}\Dellsq{\ell}(k) \,\frac{\p\ln(k^{-2}\Dellsq{\ell}(k))}{\p\ln k} \notag\\
&\ph{\int\der s}
\equiv \frac13 \int_{k_{\rm min}}^{k_{\rm max}} \der\ln k\,k^{-2}\Dellsq{\ell}(k) \, n^{(\ell)}_{\rm v}(k) \notag\\
&\ph{\int\der s}
\equiv \nbarv{\ell} \,\Sigell{\ell}\,,
\label{eq:nbarv-def}
\end{align}
where the last but one equality introduces the local logarithmic slope $n^{(\ell)}_{\rm v}(k)$ of the linear velocity power spectrum and the last line defines its averaged value \nbarv{\ell}. 

Lastly, we note that the terms multiplying \DaAP\ in the second equality in \eqn{eq:Sig2ell-obs-temp2} are superficially similar to the first term, except that terms like $\xiell{\ell}\Cal{J}_\ell$ are replaced with $\xiell{\ell^\prime}\Cal{J}_\ell$. Writing $\xiell{\ell}(s)$ as the Hankel transform \eqref{eq:Hankel}, the resulting expression can be written as a triple integral over $\int\der k$ (defining $\Cal{J}_\ell$), $\int\der k^\prime$ (defining $\xiell{\ell^\prime}(s)$ as a Hankel transform) and $\int\der s$ (which already appears). Of these, the last takes the form 
\beq
\int_0^\infty \der s\,s^2\,j_{\ell}(ks)\, j_{\ell^\prime}(k^\prime s)\,.
\label{eq:sphBessel-integral}
\eeq
This integral can be written in closed form as an infinite sum over $\ell^{\prime\prime}$ of associated Legendre functions $Q^1_{\ell^{\prime\prime}}(y)$ of the second kind, with argument $y=(k^2+k^{\prime2})/(2kk^\prime)$, weighted by combinations of Wigner $3j$ and $6j$ symbols involving $\ell$ and $\ell^\prime$ (see equation~2.8 of \cite{mehrem11}). It does not appear possible, however, to write the subsequent integrals over $k^\prime$ and $k$ in a form that can be easily approximated in the Zel'dovich smearing approximation of \citetalias{ps26a}. Since the impact of \DaAP\ is expected to be small to begin with, we instead approximate the terms in \eqn{eq:Sig2ell-obs-temp2} using
\begin{align}
\sum_{\ell^\prime} \xiell{\ell^\prime}\,\Cal{J}_\ell\,\Cal{M}_{\ell\ell^\prime} &\longrightarrow \sum_{\ell^\prime} \xiell{\ell^\prime}\,\Cal{J}_{\ell^\prime}\,\Cal{M}_{\ell\ell^\prime} \,, \notag\\
\sum_{\ell^\prime} \xiell{\ell^\prime}\,j_\ell\,\Cal{M}_{\ell\ell^\prime} &\longrightarrow \sum_{\ell^\prime} \xiell{\ell^\prime}\,j_{\ell^\prime}\,\Cal{M}_{\ell\ell^\prime} \,,
\label{eq:approxsum}
\end{align}
for some matrix $\Cal{M}_{\ell\ell^\prime}$. This allows us to write \emph{all} the resulting expressions in terms of over \Sigell{\ell^\prime} and \nbarv{\ell^\prime}, leading to 
\begin{align}
\frac{\Sigellobs{\ell}}{D_{\rm Vf}^2} &= \frac{\Sigell{\ell}}{D_{\rm V}^2}
+ \DaAP\sum_{\ell^\prime}\left(\Cal{A}_{\ell\ell^\prime} - \frac23(2\ell+1)(2+\nbarv{\ell^\prime})\,\Cal{C}_{\ell\ell^\prime}\right)  \frac{\Sigell{\ell^{\prime}}}{D_{\rm V}^2} \,.
\label{eq:Sig2ell-obs-interm}
\end{align}
We further discard the term involving \nbarv{\ell} since these factors are typically smaller than unity, at least for the monopole and quadrupole which contribute the largest signal (e.g., for the \texttt{c000} cosmology, we have $\nbarv{\ell}\simeq\{0.2,-0.34,-1.2\}$ for $\ell=0,2,4$). The difference between the factors involving $D_{\rm Vf}$ and $D_{\rm V}$ on each side is exactly a factor $\aiso^2$; this can be absorbed into the model for \Sigell{\ell} by declaring that the model actually predicts $\Sigell{\ell}\times\aiso^2$. In practice, this is accomplished by comparing constraints on the model parameter \sigv\ with the ground truth value of $\sigma_{\rm v,true}\times\aiso$, which also makes this treatment consistent with what is done for physical scales inferred from the basis coefficients $\{w_m\}$.

The final result for \Sigellobs{\ell} is then given by \eqn{eq:Sig2ell-obs}, which is in a form that can be directly predicted in the Zel'dovich smearing approximation. We test the approximations leading to the final expression in \eqn{eq:Sig2ell-obs} by direct comparison with measurements in simulations.

\begin{table}
\centering
\begin{tabular}{ccc}
\hline\hline
Parameter priors \\
\hline\hline
Parameter & \desi\ & \euclid\ \\
\hline 
$\beta$ & $\Cal{U}\left([-1,1]\right)$ & $\Cal{U}\left([-1,1]\right)$ \\
\sigv\ & $\Cal{U}\left([0,12]\right)$ & $\Cal{U}\left([0,12]\right)$ \\
$w_0$ & $\Cal{N}\left(0.0059,0.0219\right)$ & $\Cal{N}\left(0.0025,0.0094\right)$ \\
$w_1$ & $\Cal{N}\left(-0.0120,0.0292\right)$ & $\Cal{N}\left(-0.0052,0.0125\right)$ \\
$w_2$ & $\Cal{N}\left(-0.0091,0.0086\right)$ & $\Cal{N}\left(-0.0039,0.0037\right)$ \\
$w_3$ & $\Cal{N}\left(0.0160,0.0140\right)$ & $\Cal{N}\left(0.0068,0.0060\right)$ \\
$w_4$ & $\Cal{N}\left(0.0063,0.0131\right)$ & $\Cal{N}\left(0.0027,0.0056\right)$ \\
$w_5$ & $\Cal{N}\left(0.0180,0.0237\right)$ & $\Cal{N}\left(0.0077,0.0102\right)$ \\
$w_6$ & $\Cal{N}\left(-0.0172,0.0137\right)$ & $\Cal{N}\left(-0.0074,0.0059\right)$ \\
$w_7$ & $\Cal{N}\left(0.0068,0.0335\right)$ & $\Cal{N}\left(0.0029,0.0143\right)$ \\
$w_8$ & $\Cal{N}\left(0.0183,0.0140\right)$ & $\Cal{N}\left(0.0078,0.0060\right)$ \\
$f_{\rm v}$ & $\Cal{N}\left(0.26,0.09\right)$ & $\Cal{N}\left(0.26,0.09\right)$ \\
\DaAP\ & $\Cal{N}\left(0,0.097\right)$ & $\Cal{N}\left(0,0.097\right)$ \\
$b$ & $\Cal{N}\left(2.383,0.119\right)$ & $\Cal{N}\left(1.784,0.089\right)$ \\
$B_{1\ast}$ & $\Cal{U}\left([-100,100]\right)$ & $\Cal{U}\left([-100,100]\right)$ \\
$B_{v\ast}$ & $\Cal{U}\left([-100,100]\right)$ & $\Cal{U}\left([-100,100]\right)$ \\
$\sigma$ & $\Cal{U}\left([0,12]\right)$ & $\Cal{U}\left([0,12]\right)$ \\
$A_{\rm MC}$ & $\Cal{N}\left(0,0.05\right)$ & $\Cal{N}\left(0,0.05\right)$ \\
$\qbar{2}{}$ & $\Cal{U}\left([-2,2]\right)$ & $\Cal{U}\left([-2,2]\right)$ \\
\hline\hline
Physical priors \\
\hline\hline
smearing & $\sigma \geq \sqrt{2}\,\sigv$ & \\
BAO feature & $30 \leq r_{\rm peak}, r_{\rm dip},r_{\rm ZC} \leq 150$ & \\
\hline\hline
\end{tabular}
\caption{Summary of priors on all sampled parameters, along with additional physical priors imposed during likelihood evaluation. The notation $\Cal{U}\left([a,b]\right)$ denotes a uniform prior on parameter $\theta$ in the range $a\leq\theta\leq b$, while $\Cal{N}\left(\mu,\sigma_\mu\right)$ indicates a Gaussian prior on $\theta$ with mean $\mu$ and standard deviation $\sigma_\mu$. The values relevant for the length scales $\sigv,\sigma,r_{\rm peak},r_{\rm dip},r_{\rm ZC}$ are quoted in units of \Mpch, where $h=h_{\rm fid}=0.6737$ (section~\ref{subsec:cosmologies}). 
The difference between the $(\mu,\sigma_\mu)$ values for the basis coefficients $\{w_m\}$ between the \desi\ and \euclid\ cases is due to different values of the mean bias $b$ and fiducial growth factor at the respective redshift. The physical priors are common to both configurations.}
\label{tab:priors}
\end{table}

\section{Priors}
\label{app:priors}
We impose priors on all our parameters as described by \citetalias{ps26a}, additionally including priors on \Daiso\ and \DaAP. We briefly describe our choices below.

For the parameters $\left\{\{w_m\}, \Daiso,\DaAP, f_{\rm v}\right\}$, we use a `weak $\Lambda$CDM' prior as discussed by \citetalias{ps26a}. This is based on the distribution of $\Lambda$CDM parameter vectors used by \citetalias{ps25a} in their `stringent' test of the \biseq\ basis. This distribution is a Latin hypercube in the parameters $\{\Omega_{\rm m},\Omega_{\rm b},h,n_{\rm s},A_{\rm s},\Omega_{\rm k},w_{\rm DE,0}\}$ centered on the fiducial cosmology (section~\ref{subsec:cosmologies}) with ranges being $\pm5\%$ of the fiducial values for all parameters, except the dark energy equation of state $w_{\rm DE,0}$ for which the range is $\pm10\%$, and spatial curvature $\Omega_{\rm k}$ for which the range is $\pm0.05$. These Latin hypercube ranges are at least $2\times$ broader than Planck 2018 \cite{Planck18-VI-cosmoparam} constraints on all parameters at $68\%$ confidence, and are in fact $>6\times$ broader for several parameters.

The `weak $\Lambda$CDM' prior on $\left\{\{w_m\},\DaAP,f_{\rm v}\right\}$ is generated by estimating these parameters on the $\Lambda$CDM hypercube above. At this stage, the distribution is similar to the `strong $\Lambda$CDM prior' discussed in section~\ref{sec:conclude}, in that it largely preserves $\Lambda$CDM-induced correlations in the agnostic space. The prior on each model-agnostic parameter is then defined as an independent Gaussian centered on the mean of the evaluations and having a width of $10\times$ the corresponding standard deviation, thus also largely destroying any correlations induced by the $\Lambda$CDM hypothesis. In addition to not strictly obeying $\Lambda$CDM constraints, the resulting prior distribution represents regions at least $\gtrsim20\times$ broader (and often $\geq60\times$ broader) than the Planck 2018 marginal constraints at $68\%$ confidence.

The priors on all other parameters are the same as described by \citetalias{ps26a}. The most important of these is a $5\%$ Gaussian prior on the scale independent linear Eulerian bias $b$, centered on its fiducial value. In the present work, we estimate this fiducial value by fitting a constant to the ratio $\hat P(k)/P_{\rm lin}(k)$ for $0.01\leq k/(\hMpc)\leq0.06$, where $\hat P(k)$ is the real space tracer power spectrum measured in the 25 \abacussummit\ boxes (\emph{right panels} of Figs.~\ref{fig:pwise-desiLRG2} and~\ref{fig:pwise-euclidELG}). In observed samples in the DESI and Euclid surveys, we expect a constraint on $b$ to be available using weak lensing measurements (e.g., \cite{semenaite+26}). 

The other priors worth noting are those on the smearing scales $\{\sigma,\sigv\}$, each of which is uniform in the range $[0,12]\,\Mpch$. The lower limit is by definition, while the upper limit is a technical requirement imposed by the fact that the \biseq\ basis was calibrated by \citetalias{ps25a} under the assumption that any smearing scale would not exceed $\sim12\Mpch$. This choice does affect our analysis, especially for the \euclid\ sample (see section~\ref{subsec:sdbmc-constraints}). In principle, the \biseq\ basis $\{b_m(r)\}$ could be recalibrated to extend over a wider range of separations $r$, so as to allow for a wider range of smearing scales; for simplicity, we have chosen not to do so in this work. 

The full list of priors for the \desi\ and \euclid\ samples is summarized in Table~\ref{tab:priors}. 

\section{Error covariance}
\label{app:covariance}
We assume a Gaussian likelihood for our entire data set $\mathbf{z}=\{z_i\}$, where $1\leq i\leq 93$ indexes the 3 values of \Sigellsim{\ell} followed by $30$ values each of \xiellsim{\ell}. The likelihood is therefore completely specified by the error covariance matrix $C_{ij}$.

The diagonal error $\sigma_i=\sqrt{C_{ii}}$ on each measured $z_i$ is estimated as the standard deviation of $z_i$ across the 25 realizations of the \abacussummit\ \base\ cosmology:
\beq
\sigma_i^2 = \frac{1}{N-1} \,\sum_{\alpha=1}^{N}\left(z_i^{(\alpha)} -\bar z_i\right)^2\,;\quad \bar z_i \equiv \frac1N\,\sum_{\alpha=1}^N z_i^{(\alpha)}\,,
\eeq
where $N=25$ and $z_i^{(\alpha)}$ is the estimate of $z_i$ in realization $\alpha$.

\begin{figure}
\centering
\includegraphics[width=0.44\linewidth]{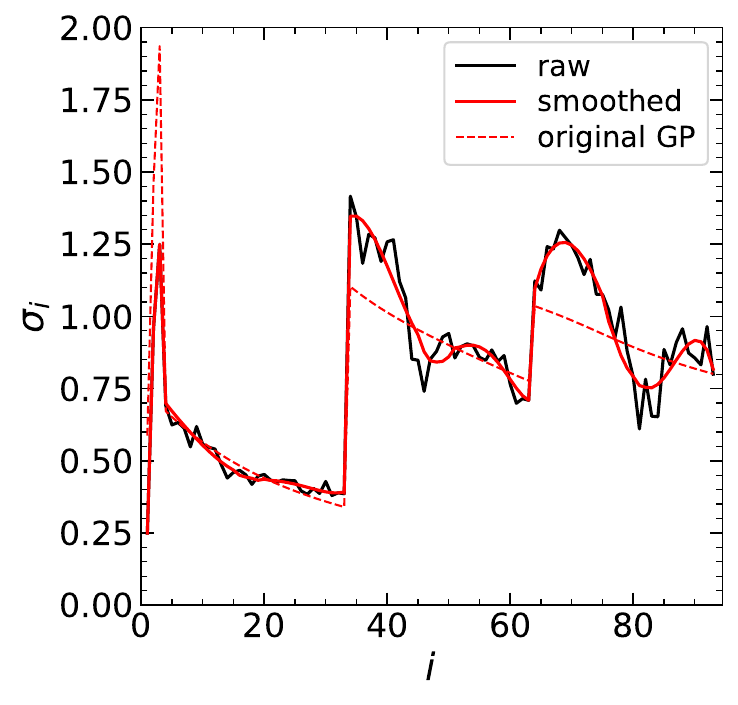}
\includegraphics[width=0.55\linewidth]{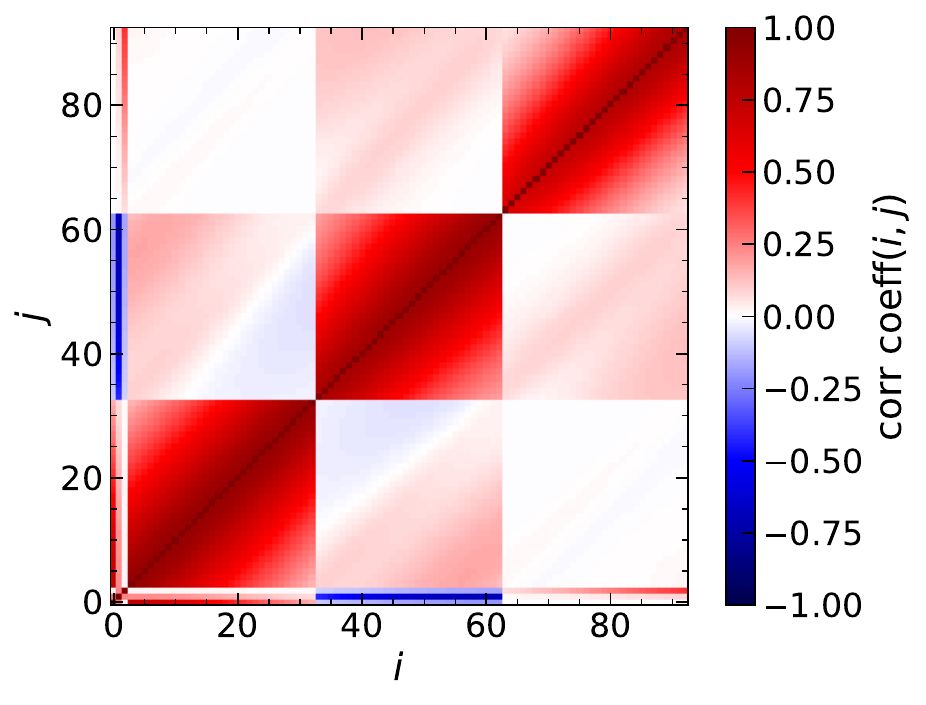}
\caption{Rescaled Gauss-Poisson (GP) covariance matrix for the \desi\ configuration. The structure of the matrix for the \euclid\ configuration is very similar. \emph{(Left panel):} Diagonal errors $\sigma_i$ on each quantity $z_i$ labelled by $1\leq i\leq 93$ (see Appendix~\ref{app:covariance}). The solid black curve shows the raw estimate from 25 realizations of the \abacussummit\ baseline \base\ cosmology, the solid red curve shows the smoothed version of the raw measurements, used as the target for the final diagonal errors and, for reference, the dashed red curve shows the original GP estimate of $\sigma_i$. For clarity, we have further scaled all 2pcf errors ($i\geq 4$) by a factor $10^3$. \emph{(Right panel):} Correlation matrix as prescribed by the GP approximation. This is used as-is in our analysis.}
\label{fig:GPcov}
\end{figure}

An error covariance $C^{\rm (GP)}_{ij}$ is first estimated using the Gauss-Poisson (GP) approximation (e.g., \cite{grieb2016}) as described by \citetalias{ps26a} (see their Appendix C). The rows and columns of $C^{\rm (GP)}_{ij}$ are then scaled so as to replace the marginal error $\sqrt{C^{\rm (GP)}_{ii}}$ on quantity $i$ with $\sigma_i$, while preserving the correlation structure, to obtain the final covariance matrix $C_{ij}$
\beq
C_{ij} = \sigma_i\,\sigma_j\,\frac{C^{\rm (GP)}_{ij}}{\sqrt{C^{\rm (GP)}_{ii}\,C^{\rm (GP)}_{jj}}}\,.
\eeq
In practice, we apply a Savitzky-Golay smoothing filter to the rescaling factor $\sigma_i/\sqrt{C_{ii}^{\rm (GP)}}$ in order to mitigate noise due to the finite number of realisations. Fig.~\ref{fig:GPcov} shows the diagonal entries \emph{(left panel)} and correlation matrix \emph{(right panel)} of the resulting covariance matrix for the \desi\ configuration. The structure of the matrix for the \euclid\ configuration is very similar, so we do not display it. Finally, we modify the 2pcf measurements and error covariance as described by \citetalias{ps26a} (their equation C.8) so as to eliminate one of the nuisance parameters in the Zel'dovich smearing approximation. The dimensionality of resulting data vector and its error covariance is $3+30+29+29=91$.

\begin{figure}
\centering
\includegraphics[width=\textwidth]{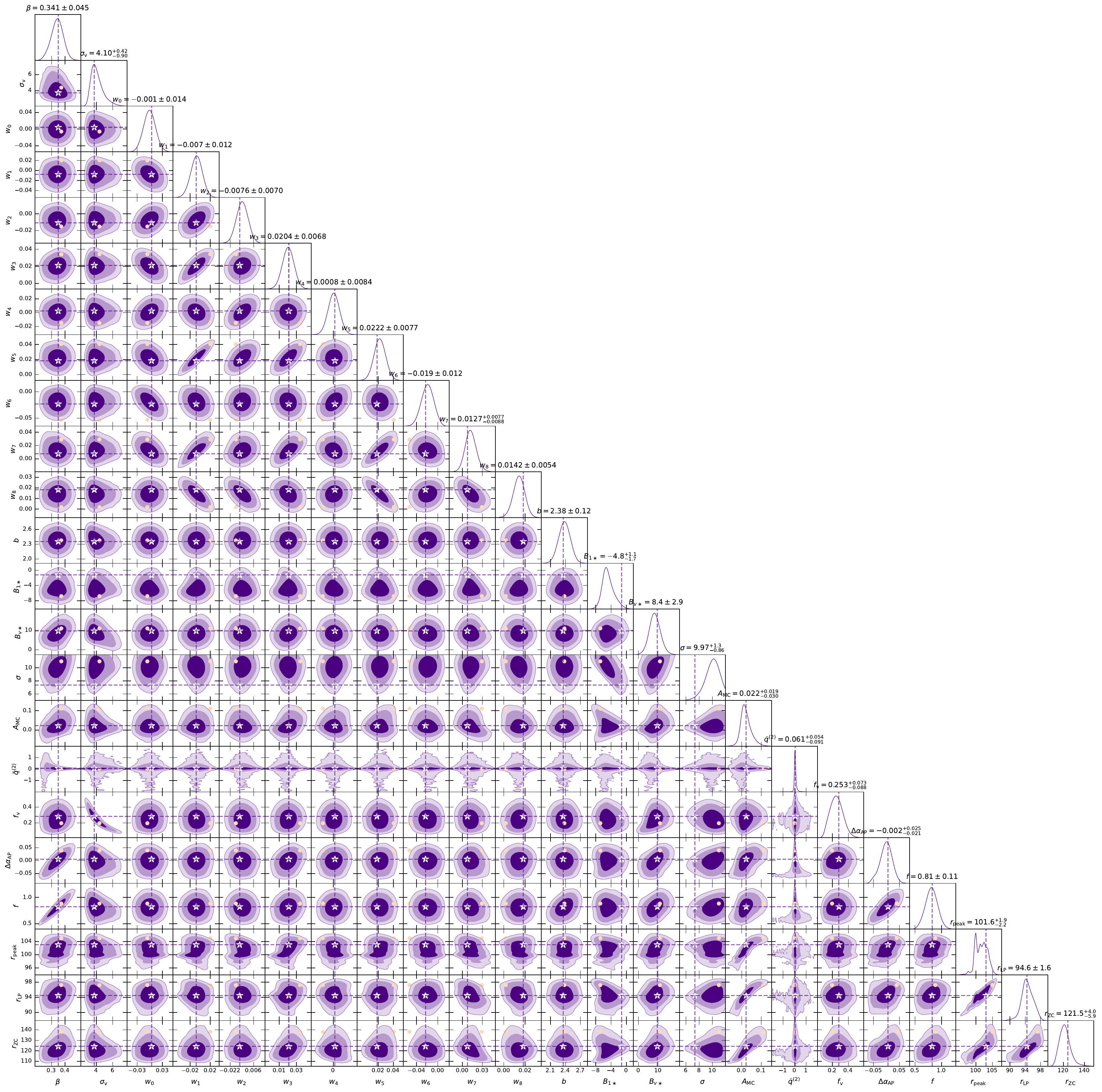}
\caption{Joint constraints on all model parameters for the inference exercise with the \desi\ sample. Note that $\{f,r_{\rm peak},r_{\rm LP},r_{\rm ZC}\}$ are derived parameters. Subsets of these constraints are displayed in Figs.~\ref{fig:cosmofit-desiLRG2} and~\ref{fig:cosmo-sdbmc-desiLRG2}. See text for a discussion.}
\label{fig:contours-desiLRG2}
\end{figure}

The calculation of $C^{\rm (GP)}_{ij}$ needs a specification of values of the \emph{sdbmc} parameters $B_{1\ast}$, $B_{v\ast}$, $\sigma$ and $A_{\rm MC}$. Since these are not known \emph{a priori}, we follow the iterative approach of \citetalias{ps25b}. We first estimate $C^{\rm (GP)}_{ij}$ by setting all these parameters to their values in the \emph{no sdbmc} model ($\{B_{1\ast},B_{v\ast},A_{\rm MC}\}\to 0$ and $\sigma\to\sqrt{2}\sigv$) and perform a `zeroth' iteration of parameter inference. The resulting best fit (MAP) values of these parameters are then used to update the estimate of $C^{\rm (GP)}_{ij}$, using which we perform a `first' iteration of inference. The best fitting parameters from this first iteration are now expected to be relatively stable. We use these to again update the estimate of $C^{\rm (GP)}_{ij}$ and perform a second iteration of inference, indeed finding shifts of less than $1\sigma$ in all parameters as compared with the previous iteration.

\begin{figure}
\centering
\includegraphics[width=\textwidth]{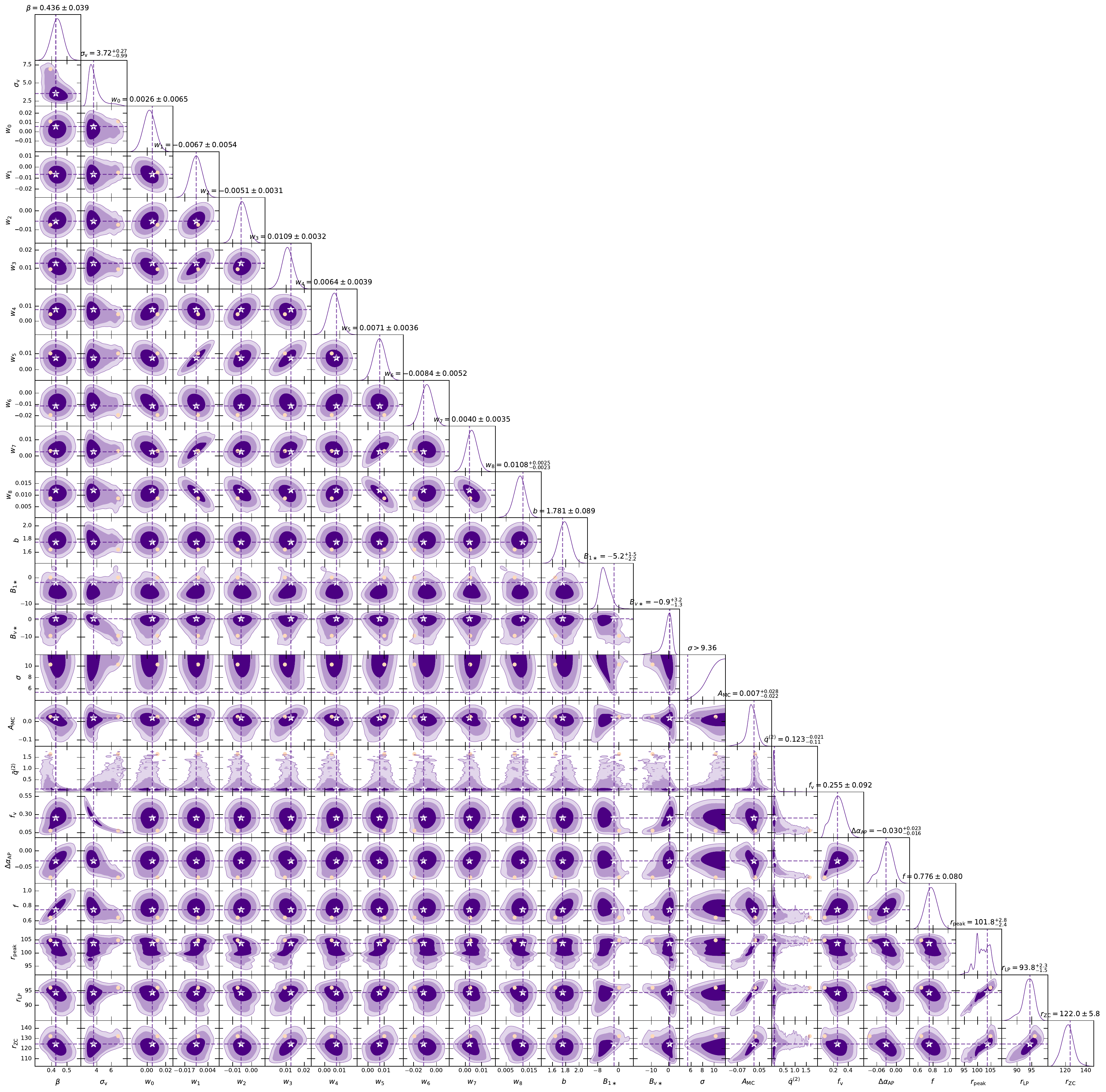}
\caption{Same as Fig.~\ref{fig:contours-desiLRG2}, for the \euclid\ sample.}
\label{fig:contours-euclidELG}
\end{figure}

Unlike \citetalias{ps25b}, but similarly to  \citetalias{ps26a}, in addition to the \emph{sdbmc} parameters $\{B_{1\ast},B_{v\ast},\sigma\}$ we also include a non-zero value of $A_{\rm MC}$ (taken from the previous iterations) when calculating the covariance matrix in iterations 1 and 2 above. We do this by approximating the mode coupling contribution to the anisotropic power spectrum as a Legendre-weighted sum over its multipoles $\ell=0,2,4$.  We also performed some tests where, similarly to \citetalias{ps25b} and \citetalias{ps26a}, we set $A_{\rm MC}\to0$ when estimating each iteration's covariance matrix. This leads to only minor effects on the final best fitting parameter values. The goodness-of-fit, however, is typically lower when including a non-zero $A_{\rm MC}$. Since the $\chi^2/{\rm d.o.f.} \sim 0.71 < 1$ for the \desi\ sample (see Fig.~\ref{fig:cosmofit-desiLRG2}), it is possible that the correlation structure predicted by the combination of the Gauss-Poisson and Zel'dovich smearing approximations is mis-estimated and slightly overpredicts the degree of correlation between the data points. Although this doesn't affect the conclusions of this work, we note that any future analyses of real data using our model-agnostic framework would benefit from using more accurately calibrated covariance matrices for the relevant data sets.

\begin{figure}
\centering
\includegraphics[width=0.9\textwidth]{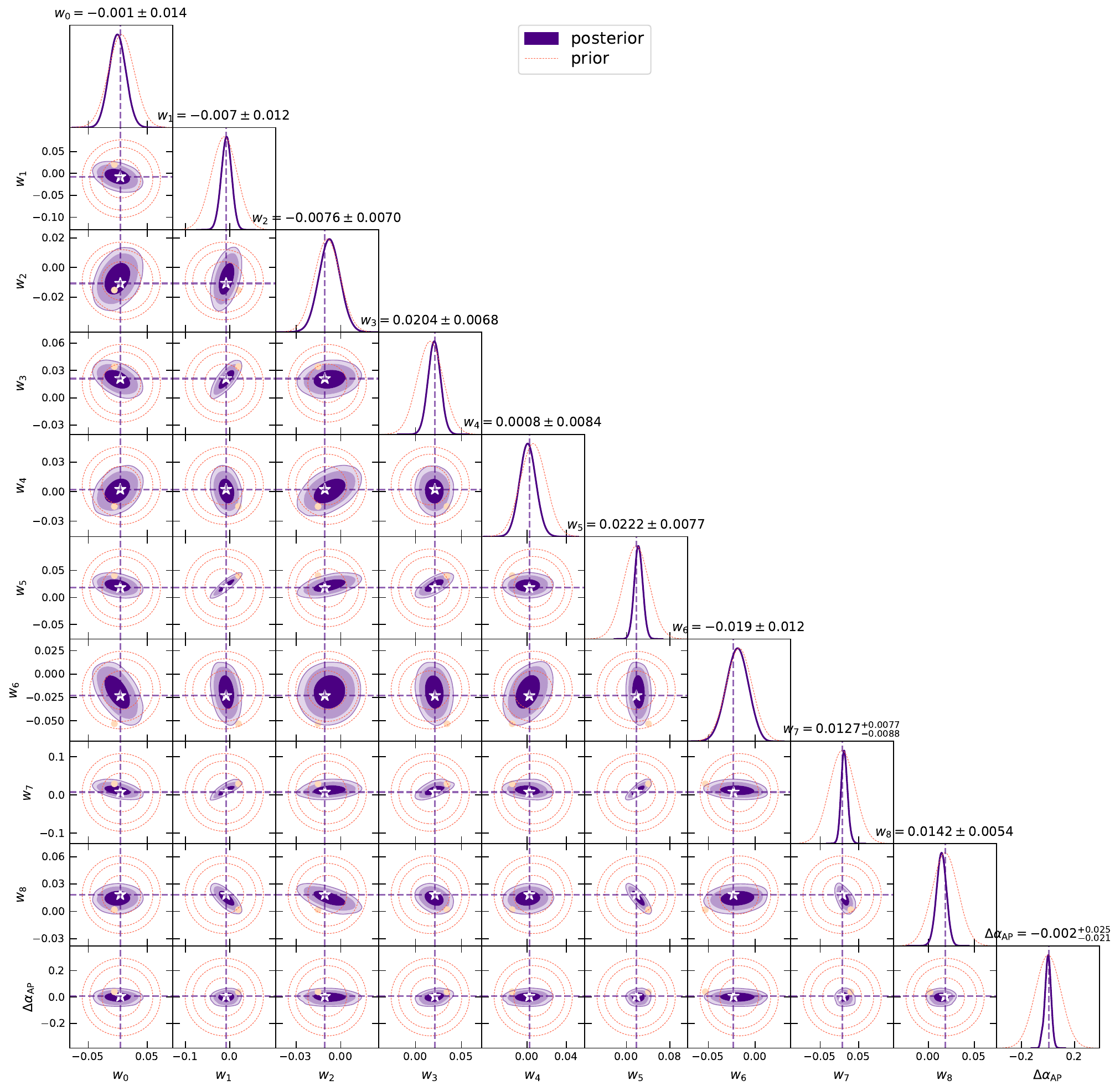}
\caption{Comparison of posterior (blue solid) and prior (red dashed) distributions for the cosmological parameters $\left\{\{w_m\},\DaAP\right\}$ for the \desi\ sample. We see that $w_1,w_3,w_5,w_7,w_8$ and \DaAP\ are informatively constrained by the data, while the remaining parameters, especially $w_6$, are largely constrained only by the priors. See text for a discussion.}
\label{fig:contours-cosmoprior-desiLRG2}
\end{figure}

Throughout, we report the results of the second iteration described above, including in Fig.~\ref{fig:GPcov}, and provide the scaled, smoothed GP estimate for each iteration of each configuration in the public repository mentioned in the `Data Availability' statement.

\section{Detailed posterior distributions}
\label{app:allparams}

Following \citetalias{ps26a}, for the various summaries presented in the main text, we compress our constraints on the $9$ basis coefficients $\{w_m\}$ into estimates of the linear point $r_{\rm LP}$ and zero crossing $r_{\rm ZC}$ of the linear theory 2pcf $\xi_{\rm lin}(r)$, to which we append an estimate of the growth rate $f=\beta\,b$.

\begin{figure}
\centering
\includegraphics[width=0.9\textwidth]{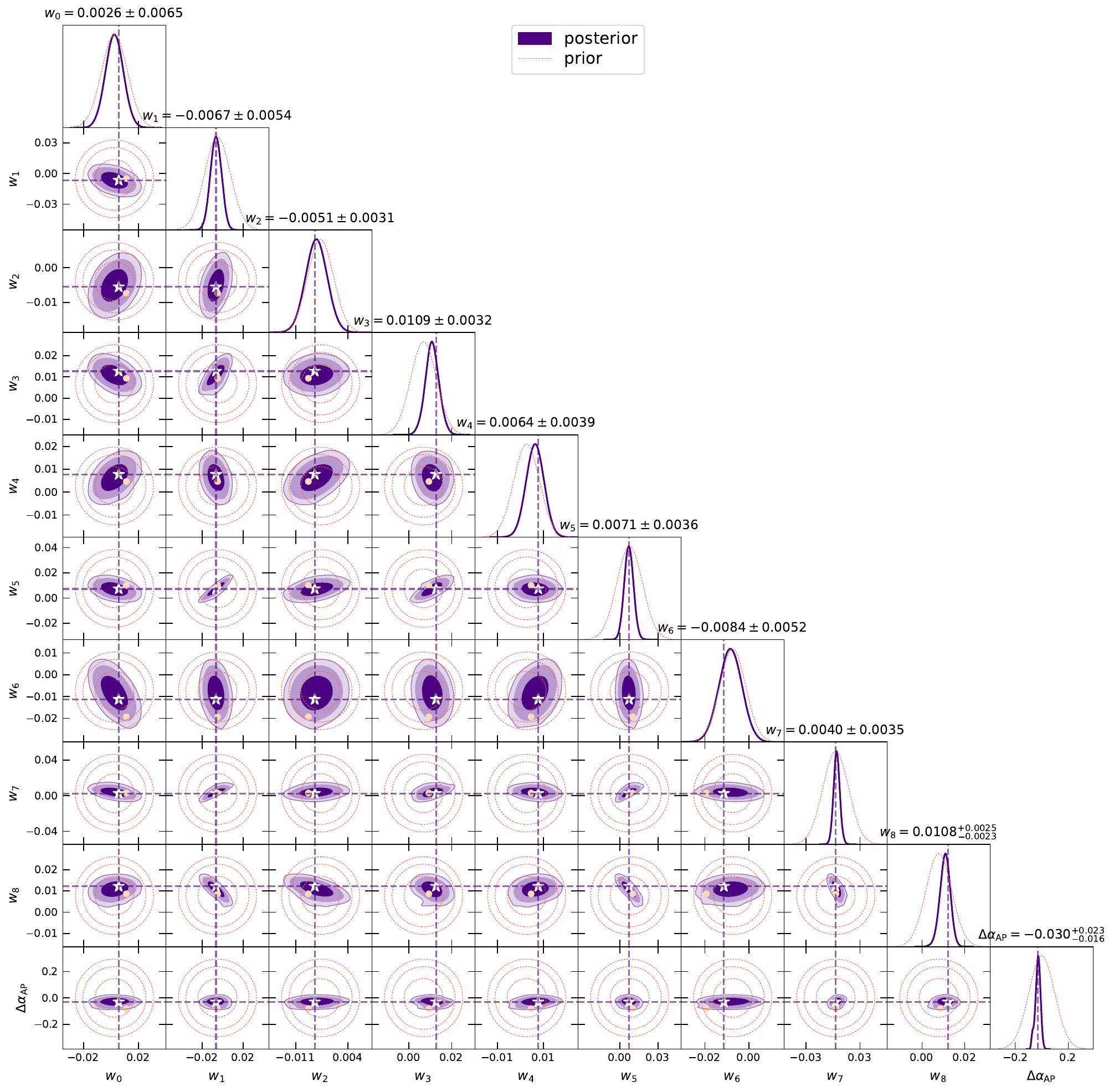}
\caption{Same as Fig.~\ref{fig:contours-cosmoprior-desiLRG2}, for the \euclid\ sample.}
\label{fig:contours-cosmoprior-euclidELG}
\end{figure}

The linear point is defined as the average of the peak $r_{\rm peak}$ and dip $r_{\rm dip}$ of $\xi_{\rm lin}(r)$,
\beq
r_{\rm LP} \equiv \left(r_{\rm peak} + r_{\rm dip}\right)/2\,.
\eeq
The details of the numerical evaluation of these scales can be found in Appendix~F of \citetalias{ps26a}, and the relevant code is included in the publicly available repository mentioned in the `Data Availability' statement above.

Figs.~\ref{fig:contours-desiLRG2} and~\ref{fig:contours-euclidELG} show the full set of pairwise posterior distributions from the MCMC analyses described in the main text, for $20$ sampled and $4$ derived parameters.

Figs.~\ref{fig:contours-cosmoprior-desiLRG2} and~\ref{fig:contours-cosmoprior-euclidELG} compare the posteriors for the cosmological parameters $\{w_m\},\DaAP$ with their priors. The results are discussed in section~\ref{subsec:constraining-power}.

\end{document}